\documentclass[review,3p,10pt,times]{./elsarticle}

\usepackage{amssymb,amsmath,amsfonts}
\usepackage{geometry}
\usepackage{algorithmic}
\usepackage{algorithm}
\usepackage{array}
\usepackage{caption}
\usepackage[caption=false,font=footnotesize,labelfont=rm,textfont=rm]{subfig}
\usepackage{textcomp}
\usepackage{stfloats}
\usepackage{url}
\usepackage{verbatim}
\usepackage{graphicx}
\usepackage{multirow}
\usepackage{arydshln}
\usepackage{bm}

\biboptions{sort&compress}
\begin{document}

\begin{frontmatter}



\title{Efficient Wideband DoA Estimation with a Robust Iterative Method for Uniform Circular Arrays\tnoteref{t1}}

\tnotetext[t1]{Part of this work \cite{ref1} was presented at the IEEE International Conference on Communications (ICC), 2023.}

\author[1]{Xiaorui Ding}
\ead{dingxr@bupt.edu.cn}
\author[1]{Wenbo Xu\corref{cor1}}
\ead{xuwb@bupt.edu.cn}
\author[2]{Yue Wang\corref{cor1}}
\ead{ywang182@gsu.edu}

\cortext[cor1]{Corresponding author}

\affiliation[1]{organization={Key Lab of Universal Wireless Communications, Ministry of Education, Beijing University of Posts and Telecommunications},
            city={Beijing},
            postcode={100876},
            country={China}}
\affiliation[2]{organization={Department of Computer Science, Georgia State University},
            city={Atlanta, GA},
            postcode={30303},
            country={USA}}

\begin{abstract}
Direction-of-arrival (DoA) is a critical parameter in wireless channel estimation. With the ever-increasing requirement of high data rate and ubiquitous devices in wireless communication systems, effective wideband DoA estimation is desirable. In this paper, an iterative coherent signal-subspace method including three main steps in each iteration is proposed for wideband two-dimensional (2D) DoA estimation with a uniform circular array. The first step selects partial frequency points for the subsequent focusing process. The second step performs the focusing process, where the angle intervals are designed to generate focusing matrices with robustness, and the signal-subspaces at the selected frequency points are focused into a reference frequency. The third step estimates DoAs with the multiple signal classification (MUSIC) algorithm, where the range of the MUSIC spatial spectrum is constrained by the aforementioned angle intervals. The key parameters of the proposed method in the current iteration are adjusted based on the estimation results in the previous iterations. Besides, the Cramér-Rao bound of the investigated scenario of DoA estimation is derived as a performance benchmark, based on which the guidelines for practical application are provided. The simulation results indicate the proposed method enjoys better estimation performance and preferable efficiency when compared with the benchmark methods.
\end{abstract}


\end{frontmatter}

\section{Introduction}

As an important part of the channel estimation in multi-antenna wireless communication systems, direction-of-arrival (DoA) estimation has been widely studied. Among different types of antenna arrays, the uniform circular array (UCA) is popular due to its capability of estimating two-dimensional (2D) DoA with unambiguous azimuth \cite{ref2}. Such benefit attracts many scholars to investigate the DoA estimation with UCAs, and the effective methods, e.g., the maximum-likelihood (ML) method \cite{ref3}, the multiple signal classification (MUSIC) algorithm \cite{ref4}, the estimation of signal parameters via rotational invariance techniques \cite{ref5} and some deep-learning-based methods \cite{ref6,ref7,ref8} are developed for estimating DoAs with a UCA. These methods achieve different trade-offs between DoA estimation performance and computational complexity, thereby satisfying the demands of various scenarios.

MUSIC algorithm is the most popular one among these methods due to its excellent performance and reasonable complexity \cite{ref9,ref10,ref11}. It utilizes the eigenvalue decomposition (EVD) of the signal covariance matrix to construct the signal-subspace and the noise-subspace. Then, the DoAs of multiple signals are obtained through peak-searching in a spatial spectrum formed by the orthogonality of the two subspaces. Furthermore, when the arriving signals are uncorrelated and sampled with enough snapshots, the performance of MUSIC algorithm is close to that of ML method \cite{ref9} which has perfect performance but suffers from heavy complexity burden.

Nonetheless, MUSIC algorithm is only suitable for the narrowband DoA estimation. Fortunately, some practical methods are investigated to estimate the DoAs of wideband signals that are common in wireless communication. For instance, incoherent signal-subspace method (ISM) transforms the wideband signal-subspace into narrowband ones by Fourier transform \cite{ref12,ref13,ref14}. Then, ISM respectively performs narrowband DoA estimation method on each narrowband signal-subspaces and finally synthesizes the results. Alternatively, Jacobi-Anger app-roximation transforms all narrowband manifold matrices into a manifold matrix of a virtual uniform linear array, and then narrowband methods can be used to estimate DoAs \cite{ref15,ref16,ref17}.

Considering the multipath effect in wireless communication, the wideband signals received by antenna arrays may be from different paths of the same signal with small time intervals, which are approximately coherent and disable the aforementioned classical methods to estimate the DoAs accurately. Considering such problem, \cite{ref18} proposed the coherent signal-subspace method (CSM), which is based on the coarse pre-estimated DoAs and shows preferable performance for approximate coherent signals. CSM refines the pre-estimated DoAs by MUSIC algorithm after focusing signal-subspaces at each frequency point into a reference frequency by focusing matrices. However, its performance is affected by the exactness of focusing matrices and the accuracy of pre-estimated DoAs.

To improve the performance of CSM, some methods have been proposed to construct more exact focusing matrices, including rotational signal-subspace (RSS) \cite{ref19}, signal-subspace transformation \cite{ref20}, two-sided correlation transformation \cite{ref21}, etc. Apart from these methods, some works adopt other strategies to further improve the performance of CSM. \cite{ref19} presented the iteration mechanism and added extra angles for focusing, where the angles participated in focusing process are called focusing angles. Additionally, \cite{ref22} proposed the robust CSM (R-CSM) that samples focusing angles from a large robustness interval and iteratively estimates DoAs with CSM, so as to reduce the effect of unfavorable pre-estimated DoAs on the estimation performance. \cite{ref23} investigated the iterative CSM based on two-dimensional (2D) discrete Fourier transform (I-2D-CSM) to further improve the performance, which optimizes the robustness interval of R-CSM by setting a minimum radius for it. However, these methods are only designed for one-dimensional (1D) DoA estimation, and the robustness intervals of R-CSM and I-2D-CSM are proposed without refinement, which can be further improved.

Besides, Cramér-Rao bound (CRB) is a typical benchmark of DoA estimation performance, which indicates the lower bound of the variance of the estimated DoA \cite{ref9}. Many works have derived the CRB of the DoA estimation in various scenarios. For instance, \cite{ref24} derived the CRB of 1D DoA estimation of narrowband signals with multiple snapshots and a UCA, \cite{ref25} derived the CRB of 2D DoA estimation of wideband signals with a single snapshot and a UCA, and \cite{ref26} derived the CRB of 2D DoA estimation of wideband uncorrelated signals with multiple snapshot and a planar array. However, none of the existing CRBs can be applied to the 2D DoA estimation methods which are suitable for the correlated non-zero wideband signals received by a UCA with multiple snapshots, e.g., CSM.

To achieve better performance than the aforementioned CSM methods \cite{ref19,ref22,ref23} without introducing high computational complexity, a wideband 2D DoA estimation method named the robust iterative partial-focusing CSM (RIPF-CSM) is proposed in this paper. Then, the CRB is derived to give guidance for corresponding wideband 2D DoA estimation. The main contributions of this paper are summarized as follows.

\begin{enumerate}[~~~~~~1)]
\item{An efficient iterative 2D DoA estimation method is proposed for UCAs, called RIPF-CSM. In the proposed method, the robustness intervals with less redundant focusing angles are designed to improve the focusing performance and reduce computational complexity. In order to further accelerate the estimation procedure, the range of MUSIC spatial spectrum is shrunk and the number of candidate frequency points in focusing process is reduced. During the iterations, the parameters of the above designs are adjusted based on the estimated DoAs in the previous iterations to improve the reliability.}
\item{The computational complexity of RIPF-CSM is analysed and compared with the 2D DoA estimation version of the benchmark methods that are extended from the original 1D versions. Additionally, a parameter constraint is derived, which guarantees the computational complexity of RIPF-CSM lower than that of any benchmark method in a single iteration, and it can be easily satisfied in common scenarios.}
\item{The CRB of 2D DoA estimation with a UCA is derived while considering arbitrary multiple non-zero wideband signal sources and multiple snapshots. Based on the theoretical analysis of the derived CRB, some effective guidelines are provided for the practical application of the wideband 2D DoA estimation with a UCA.}
\end{enumerate}

The rest of this paper is organized as follows. In Section 2, this paper introduces the signal model for wideband 2D DoA estimation with a UCA. Then, the RIPF-CSM is proposed and its computational complexity is discussed in Section 3. In Section 4, the CRB of DoA estimation is derived, based on which the guidelines for practical application are provided. The simulation results are presented in Section 5, followed by conclusions in Section 6.

\textit{Notations}: In this paper, $e$ denotes the natural constant. $j=\sqrt{-1}$ stands for the imaginary unit. $\mathbf{1}_M$ and $\mathbf{I}_M$ respectively represent $M\times M$ all-ones matrix and $M\times M$ identity matrix. ${{\mathbb{R}}^{M\times N}}$ and ${{\mathbb{C}}^{M\times N}}$ denote the sets of $M\times N$ real and complex matrices, respectively. ${[\cdot]^{-1}}$, $[\cdot]^\text{T}$ and ${{[\cdot]}^{\text{H}}}$ stand for the inverse, transpose and conjugate transpose of a matrix, respectively. $\left\lvert\cdot\right\rvert$ represents the absolute value of a scalar or the number of elements in a set. $\left\lvert\left\lvert\cdot\right\rvert\right\rvert_\text{F}$ stands for the Frobenius norm of a matrix. $\mathcal{E}[\cdot ]$ denotes the expectation of a random variable. $\mathcal{P}[\cdot]$ stands for and the probability of a random variable. $\left\lceil \cdot \right\rceil$ and $\left\lfloor\cdot \right\rfloor$ represent the round-up and round-down operation, respectively. $\max[\cdot]$, $\min[\cdot]$ and $\min[\cdot\vert\mathcal{H}]$ stand for the largest number, the smallest number, and the smallest number with condition $\mathcal{H}$, respectively. $\arg\min[\cdot]$ obtains the optimal variable value that minimizes the expression in the bracket. $\operatorname{frac}[\cdot]$ denotes the fractional part of a number. $\operatorname{diag}[\cdot]$ denotes a diagonal or block diagonal matrix, where the entries in the bracket respectively stand for the diagonal elements or blocks in order. $\operatorname{trace}[\cdot ]$ denotes the trace of a matrix. $\operatorname{Re}[\cdot]$ and $\operatorname{Im}[\cdot]$ represent the real and imaginary parts of a complex number, respectively. $\odot$ stands for Hadamard product. $\mathcal{O}(N)$ represents the number of operations, which is less than $CN$ as $N$ tends to infinite, and $C$ is a positive real number.

\section{Signal Model}

\begin{figure}[!b]
\centering
\vspace{-2mm}
\includegraphics[width=2.8in]{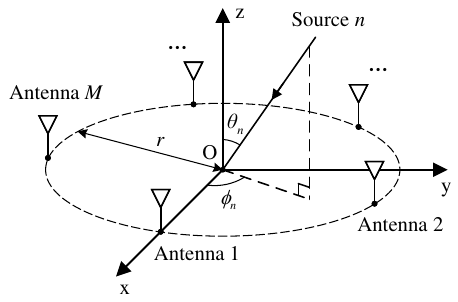}
\vspace{-2mm}
\caption{Signal receiving geometry of the UCA.}
\label{fig1}
\end{figure}

This paper considers a UCA which is equipped by $M$ omni-directional antenna elements in the x-y plane as shown in Fig. 1. Suppose there are $N$ $(N<M)$ far-field incident wide-band signals from different directions with central frequency $f_0$, and the arriving elevation and azimuth of the $n$-th signal source are respectively denoted as $\theta_n$ and $\phi_n$, where $n\in \{1,2,\dots,N\}$, ${{\theta }_{n}}\in [{{0}^{\circ }},{{90}^{\circ }}]$ and ${{\phi }_{n}}\in [{{0}^{\circ }},{{360}^{\circ }})$. The distance between adjacent antennas is set as the half of the minimum wavelength $\lambda_\text{min}$ of the received signals. Thus, the radius $r$ of the UCA is computed as $r={{\lambda }_{\text{min}}}/[4\sin(\pi/M)]$ \cite{ref4}.

With the sampling rate $f_\text{S}$ that avoids the spectrum aliasing, the output samples of UCA are collected in the matrix ${{\mathbf{X}}_{t}}\in {{\mathbb{C}}^{M\times {{K}_{t}}}}$, where ${{K}_{t}}=\left\lfloor {{t}_{0}}{{f}_{\text{S}}} \right\rfloor$ denotes the number of snapshots, and $t_0$ represents the sampling duration. The samples in the $m$-th row of ${\mathbf{X}}_{t}$ correspond to the analytic signal from the $m$-th antenna, where $m\in\{1,2,\dots,M\}$. Meanwhile, the noise in channels is supposed to be additive white Gaussian noise with zero mean and variance ${{\sigma }^{2}}$.

Since the expression of the steering vector can only apply to the narrowband signals, this paper studies the wideband signals by establishing their multiple narrow subbands instead of the original whole wideband. The $z$-th subband corresponds to the $z$-th frequency point, where $z\in\{1,2,\dots,Z\}$, $Z<K_t$, and $Z$ frequency points are derived by the $Z$-point fast Fourier transform (FFT) of the original wideband signals \cite{ref18,ref19,ref20,ref21,ref22,ref23}. Based on the Nyquist sampling theorem, the frequency point $f_z$ is calculated as:
\begin{equation}
f_z=f_\text{S}\big[(z-1)/Z+\left\lfloor f_0/f_\text{S}-0.5\right\rfloor+\eta\big]\text{,}
\end{equation}
where $\eta=\lfloor(Z\times\operatorname{frac}[f_0/f_\text{S}-0.5])/(z-1)\rfloor$. The generation of the received signal matrix at different frequency points are described as follows. First, each row of ${{\mathbf{X}}_{t}}$ is split into $K_f$ segments, where each segments has $K_f=\left\lfloor K_t/Z\right\rfloor$ elements. Second, $Z$-point FFT is respectively performed on each segment. Third, the received signal matrix ${{\mathbf{X}}_{f}}({{f}_{z}})\in {{\mathbb{C}}^{M\times {{K}_{f}}}}$ at the frequency point $f_z$ is generated by the $z$-th element of the FFT result of all the segments, where these elements are arranged in the order of their corresponding segments in ${{\mathbf{X}}_{t}}$. In this sense, $K_f$ is also known as the number of snapshots in frequency domain. Accordingly, ${\mathbf{X}}_{f}({{f}_{z}})$ can be expressed as \cite{ref18,ref19,ref20,ref21,ref22,ref23}:
\begin{equation}
\mathbf{X}_f(f_z)=\mathbf{A}(f_z,\mathbb{S})\mathbf{S}_f(f_z)+\mathbf{W}(f_z)\text{,}
\end{equation}
where $\mathbb{S}$ denotes the set including all DoAs of signal sources, $\mathbf{A}({{f}_{z}},\mathbb{S})=[\mathbf{a}({{f}_{z}},{{\theta }_{1}},{{\phi }_{1}}),\mathbf{a}({{f}_{z}},{{\theta }_{2}},{{\phi }_{2}}),\dots,\mathbf{a}({{f}_{z}},{{\theta }_{N}},{{\phi }_{N}})]$ is the manifold matrix, $\mathbf{a}({{f}_{z}},{{\theta }_{n}},{{\phi }_{n}})$ represents the steering vector of the $n$-th signal source, $n\in\{1,2,\dots,N\}$, ${{\mathbf{S}}_{f}}({{f}_{z}})\in {{\mathbb{C}}^{N\times {{K}_{f}}}}$ and $\mathbf{W}({{f}_{z}})\in {{\mathbb{C}}^{M\times {{K}_{f}}}}$ respectively denote the source signal matrix and the Gaussian noise matrix at the frequency point $f_z$. The reference position of zero phase is regarded as the center of UCA. Accordingly, the steering vector of the $n$-th signal source is expressed as:
\begin{equation}
\mathbf{a}(f_z,\theta_n,\phi_n)=\Big[e^{j\frac{2\pi rf_z}{c}\sin\theta_n\cos(\frac{2\pi\times 0}{M}-\phi_n)},e^{j\frac{2\pi rf_z}{c}\sin\theta_n\cos\left(\frac{2\pi\times 1}{M}-\phi_n\right)},\dots,e^{j\frac{2\pi rf_z}{c}\sin\theta_n\cos\left(\frac{2\pi\times(M-1)}{M}-\phi_n\right)}\Big]_\text{\normalsize ,}^\text{T}
\end{equation}
where $c$ represents the light speed. Based on the model in (2), this paper aims to estimate the 2D DoAs from the narrowband signal matrices $\mathbf{X}_f(f_z)$, $z=1,2,\dots,Z$.

\section{Wideband DoA Estimation}

In this section, the conventional CSM (C-CSM) is first described, improved from which an efficient wideband 2D DoA estimation method named RIPF-CSM is proposed. Then, the computational complexity of RIPF-CSM is analysed and compared with the benchmark methods.

\subsection{The conventional CSM}
C-CSM is a classical DoA estimation method, which refines the coarse pre-estimated DoAs acquired by some methods with low complexity \cite{ref18, ref19, ref20}, e.g., conventional beamforming method \cite{ref27}. The key step of C-CSM is the focusing process that transforms the signal-subspaces at different frequency points to the one at a reference frequency, where the signal-subspace at the frequency point $f_z$ is spanned by the steering vectors in $\mathbf{A}(f_z,\mathbb{S})$ \cite{ref9}. In this paper, $f_0$ is regarded as the reference frequency. Define the set of the pre-estimated DoAs as $\mathbb{S}_\text{p}$. Considering only the set $\mathbb{S}_\text{p}$ is known rather than the ground-truth set $\mathbb{S}$, the focusing process at the frequency point $f_z$ depends on the focusing matrix $\mathbf{B}(f_z)$ that guarantees $\mathbf{B}(f_z)\mathbf{A}(f_z,\mathbb{S}_\text{p})\approx\mathbf{A}(f_0,\mathbb{S}_\text{p})$. To obtain $\mathbf{B}(f_z)$ while balancing the exactness and the computational complexity, the RSS method is utilized in this paper, which realizes the following optimization \cite{ref19}:
\begin{equation}
\underset{\mathbf{B}(f_z)}{\arg\min}\left[\left\lvert\left\lvert\mathbf{A}(f_0,\mathbb{S}_\text{p})-\mathbf{B}(f_z)\mathbf{A}(f_z,\mathbb{S}_\text{p})\right\rvert\right\rvert_\text{F}\right]\text{.}
\end{equation}
The result of (4) is derived as \cite{ref19}:
\begin{equation}
\mathbf{B}(f_z)=\mathbf{U}_\text{R}(f_z)\mathbf{U}_\text{L}^\text{H}(f_z)\text{,}
\end{equation}
where ${{\mathbf{U}}_{\text{L}}}({{f}_{z}})$ and ${{\mathbf{U}}_{\text{R}}}({{f}_{z}})$ denote the left and right singular matrix of $\mathbf{A}({{f}_{z}},{{\mathbb{S}}_{\text{p}}}){{\mathbf{A}}^{\text{H}}}({{f}_{0}},{{\mathbb{S}}_{\text{p}}})$ from singular value decomposition (SVD), respectively. With $\mathbf{B}(f_z)$ in (5), the focused signal matrix at the frequency point $f_z$ is computed as $\mathbf{B}(f_z)\mathbf{X}_f(f_z)$. As a result, the average covariance matrix $\mathbf{R_Y}$ of the focused narrowband signals is calculated as:
\begin{equation}
\mathbf{R_Y}=\frac{1}{K_fZ^2}\sum\limits_{z=1}^{Z}[\mathbf{B}(f_z)\mathbf{X}_f(f_z)][\mathbf{B}(f_z)\mathbf{X}_f(f_z)]^\text{H}\text{.}
\end{equation}

After the focusing process, the MUSIC algorithm is used to estimate the DoAs of the incident signals, which is introduced as follows. Based on the EVD on $\mathbf{R_Y}$, the eigenvectors corresponding to the signal-subspace and the noise-subspace are distinguished based on the eigenvalues of $\mathbf{R_Y}$, while the source number $\hat{N}$ is also obtained in this process \cite{ref28}. In addition, the noise-subspace matrix ${{\mathbf{E}}_{\text{n}}}$ is formed with the eigenvectors of $\mathbf{R_Y}$ corresponding to the noise-subspace. Then, the MUSIC spatial spectrum is generated based on the orthogonality between the noise subspace and the steering vectors of different directions. The direction of the top $\hat{N}$ peak values in the MUSIC spatial spectrum are regarded as the DoA estimation results, where the value of the spectrum with the elevation $\theta\in[{{0}^{\circ }},{{90}^{\circ }}]$ and the azimuth $\phi\in[{{0}^{\circ }},{{360}^{\circ }})$ is calculated as \cite{ref4}:
\begin{equation}
\mathit{\Theta}_\text{MUSIC}(\theta,\phi)=\frac{1}{\mathbf{a}^\text{H}(f_0,\theta,\phi)\mathbf{E}_\text{n}\mathbf{E}_\text{n}^\text{H}\mathbf{a}(f_0,\theta,\phi)}\text{.}
\end{equation}
To improve the estimation performance, the aforementioned focusing process and the MUSIC algorithm are iteratively implemented in C-CSM \cite{ref19}.

However, C-CSM suffers from the following two shortcomings. First, the errors between the pre-estimated DoAs and the actual ones significantly affect the focusing performance \cite{ref19,ref22,ref23}. Second, the numerous frequency points, the calculation of the large MUSIC spatial spectrum and the peak-searching process lead to high computational overhead.

\subsection{The proposed RIPF-CSM}
\begin{figure*}[!tb]
\centering
\includegraphics[width=6.7in]{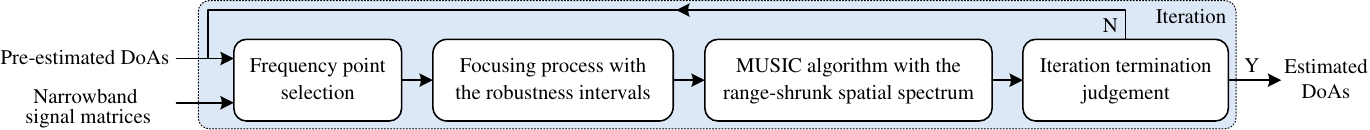}
\caption{The main structure of RIPF-CSM.}
\label{fig2}
\vspace{-2mm}
\end{figure*}

To overcome the shortcomings of C-CSM, the more efficient RIPF-CSM is proposed. Fig. 2 illustrates the main structure of RIPF-CSM, which works in an iterative manner. In this figure, the frequency point selection part randomly selects candidate frequency points for the subsequent focusing. Then, the signal-subspaces at the candidate frequency points are focused into a reference frequency with the angles in the robustness intervals during the focusing process, followed by the MUSIC algorithm which estimates the DoA of the signals with range-shrunk spatial spectrum. Finally, it is judged whether the iteration should terminate. The details are described as follows.

1) \textit{Frequency point selection}

In the first iteration, RIPF-CSM randomly selects only one frequency point, which is added into the empty focusing frequency point set $\mathbb{F}_{\text{in}}^{(1)}$, so as to minimize the computational complexity of the focusing process at the beginning. Considering the focusing matrices are obtained by following the optimization (4), the focusing error at each frequency point can be different, and thus the focusing performance may vary at different frequency points. Therefore, since the frequency point that enjoys better focusing performance cannot be figured out, it is reliable to take more frequency points into account when the DoA estimation performance is unsatisfactory. In the $i$-th iteration where $i>1$, $\mathbb{F}_{\text{in}}^{(i)}$ is generated by merging $\mathbb{F}_{\text{in}}^{(i-1)}$ and some other randomly selected frequency points, where the number of the extra selected frequency points is elaborated as follows.

Denote ${\bar{\delta }}^{(i)}$ as the average difference between the estimated DoAs in the $i$-th iteration ($i>1$) and those in the previous iteration. ${\bar{\delta }}^{(i)}$ is given as:
\begin{equation}
\bar\delta^{(i)}=
\begin{cases}
\displaystyle\frac{1}{2\hat{N}_i}\sum\limits_{n=1}^{\hat{N}_i}\lvert\hat\theta_n^{(i)}-\hat\theta_n^{(i-1)}\rvert+\lvert\hat\phi_n^{(i)}-\hat\phi_n^{(i-1)}\rvert, \text{if}~\hat{N}_{i-1}=\hat{N}_i\text{,} \\
\displaystyle\frac{1}{2\hat{N}_i}\sum\limits_{n=1}^{\hat{N}_i}\min\Big[\lvert\hat\theta_n^{(i)}-\hat\theta_{n^\prime}^{(i-1)}\rvert+\lvert\hat\phi_n^{(i)}-\hat\phi_{n^\prime}^{(i-1)}\rvert \Big\vert n^{\prime}\in\{1,2,\dots,\hat{N}_{i-1}\}\Big],~\text{if}~\hat{N}_{i-1}\ne \hat{N}_i\text{,}
\end{cases}
\end{equation}
where $\hat{N}_i$, $\hat\theta_n^{(i)}$, $\hat\phi_n^{(i)}$ respectively denote the number of estimated DoAs, the estimated elevation and azimuth of the $n$-th signal source in the $i$-th iteration. Specially, $\hat{N}_0$, $\hat\theta_n^{(0)}$, $\hat\phi_n^{(0)}$ denote the number of the pre-estimated DoAs, the elevation and azimuth of the $n$-th pre-estimated DoA, respectively. Additionally, $\overline{\theta_\text{e}}$ and $\overline{\phi_\text{e}}$ are respectively defined as the average error of elevation and azimuth of the pre-estimated DoAs, which can be easily obtained by evaluating the performance of the DoA pre-estimation method. Then, the number of frequency points that merge with $\mathbb{F}_{\text{in}}^{(i-1)}$ ($i>1$) to generate $\mathbb{F}_{\text{in}}^{(i)}$ is designed as:
\begin{equation}
\varDelta _{f}^{(i)}=\min\left[Z-\left\lvert\mathbb{F}_{\text{in}}^{(i-1)}\right\rvert,\left\lceil\left(\frac{Z}{I}+\frac{d_\theta+d_\phi}{2}\right)d_{\bar{\delta}}^{(i-1)}\right\rceil\right]\text{,}
\end{equation}
where ${{d}_{\theta }}=\overline{{{\theta }_\text{e}}}/{{1}^{\circ }}$, ${{d}_{\phi }}=\overline{{{\phi }_\text{e}}}/{{1}^{\circ }}$, $d_{{\bar{\delta }}}^{(i-1)}={{\bar{\delta }}^{(i-1)}}/{{1}^{\circ }}$, and $I$ denotes the upper limit of the iteration number. Each element in (9) are respectively explained as follows.
\vspace{-0.5mm}
\begin{itemize}
\item{\textbf{About} $\min[\cdot]$ \textbf{and} $Z-\left\lvert\mathbb{F}_{\text{in}}^{(i-1)}\right\rvert$\textbf{:} They are used to prevent $\big\lvert\mathbb{F}_{\text{in}}^{(i)}\big\rvert$ exceeding the number $Z$ of total frequency points.}\vspace{-1mm}
\item{\textbf{About} $Z/I+({{d}_{\theta }}+{{d}_{\phi }})/2$\textbf{:} Considering $Z/I$ represents the average increment of the number of candidate frequency points in each iteration, $Z/I$ is regarded as a basis of such increment. Furthermore, since worse pre-estimation performance generally requires more information in the candidate frequency points to refine, $({{d}_{\theta }}+{{d}_{\phi }})/2$ is added as a part of the aforementioned basis.}\vspace{-1mm}
\item{\textbf{About} $d_{\bar{\delta}}^{(i-1)}$\textbf{:} From (8), $\bar{\delta }^{(i)}$ as well as $d_{\bar{\delta}}^{(i-1)}$ reflects the convergence degree of the $i$-th iteration. Thus, $d_{\bar{\delta}}^{(i-1)}$ is designed as a coefficient to adjust the basis $Z/I+({{d}_{\theta }}+{{d}_{\phi }})/2$, so as to reduce unnecessary complexity while maintaining the DoA estimation performance.}\vspace{-1mm}
\item{\textbf{About} $\lceil\cdot\rceil$\textbf{:} To prevent the phenomenon that the iteration converges with inferior focusing performance, $\mathbb{F}_{\text{in}}^{(i)}$ should keep updating to avoid being trapped by unfavorable situations. Therefore, $\lceil\cdot\rceil$ in (9) is designed to set a lower bound of the number of newly added candidate frequency points to $1$.}\vspace{-0.5mm}
\end{itemize}

Set $\varDelta_{f}^{(1)}=0$ for simplicity, then the number of candidate frequency points in the $i$-th iteration ($i>1$) is expressed as:
\begin{equation}
Z_\text{in}^{(i)}=1+\sum\limits_{i^\prime=1}^i\varDelta_{f}^{(i^\prime)}\text{.}
\end{equation}

2) \textit{Focusing process with the robustness intervals}

In the focusing process of RIPF-CSM, the participated narrowband signal matrices only includes the ones at the frequency points in the set $\mathbb{F}_{\text{in}}^{(i)}$. The focusing angles of the focusing process are sampled from the designed angle intervals which are called robustness intervals. In the $i$-th iteration, the center of the $n$-th region that is constrained by the robustness intervals is set as $(\hat\theta_n^{(i-1)},\hat\phi_n^{(i-1)})$ where $n\in\{1,2,\dots,\hat{N}_{i-1}\}$, and the radii of the corresponding elevation and azimuth robustness interval are respectively denoted as $R_n^\theta(i)$ and $R_n^\phi(i)$. Compared with the radii of the existing robustness intervals \cite{ref22,ref23}, $R_n^\theta(i)$ and $R_n^\phi(i)$ are designed with the extra information including the error of the pre-estimated DoAs, the accuracy of the angles of a steering vector in noisy cases, and the estimation results in the previous iterations. The details of $R_n^\theta(i)$ and $R_n^\phi(i)$ are described as follows.

Considering the focusing angles which are close to the actual DoAs improve the focusing performance and the ones that are far away from the actual DoAs cause performance loss \cite{ref19,ref29}, it is reasonable to set the range constrained by the robustness intervals to cover actual DoAs while keeping the intervals as small as possible \cite{ref30,ref31}. To achieve such purpose, $R_n^\theta(i)$ and $R_n^\phi(i)$ are given as:
\begin{equation}
R_n^\theta(i)=\frac{\overline{\theta_\text{e}}\big[b-\cos\hat\theta_n^{(i-1)}\big]d_{{\bar{\delta }}}^{(i-1)}}{i}\text{,}
\end{equation}
\begin{equation}
R_n^\phi(i)=\frac{\overline{\phi_\text{e}}\big[b-\sin\hat\theta_n^{(i-1)}\big]d_{{\bar{\delta }}}^{(i-1)}}{i}\text{,}
\end{equation}
where $b$ denotes a constant. Specially, $d_{{\bar{\delta }}}^{(0)}$ is set to be $1$ to invalidate its effect on the radii in the first iteration. Each element in (11) and (12) are respectively explained as below.
\begin{itemize}
\item{\textbf{About} $\overline{\theta_\text{e}}$ \textbf{and} $\overline{\phi_\text{e}}$\textbf{:} Since $\overline{\theta_\text{e}}$ and $\overline{\phi_\text{e}}$ can be regarded as the approximate error of elevation and azimuth of the pre-estimated DoAs, they are respectively considered as the baselines when designing $R_{n}^{\theta }(i)$ and $R_{n}^{\phi }(i)$. In doing so, the range constrained by robustness intervals is more possibly to cover the actual DoAs, and thus such design achieves higher robustness to the error of the pre-estimated DoAs.}
\item{\textbf{About} $i$ \textbf{and} $d_{{\bar{\delta }}}^{(i-1)}$\textbf{:} Considering the estimation results become more and more accurate with iterations and $\bar{\delta}^{(i)}$ in (8) reflects the estimation accuracy of the $i$-th iteration ($i>1$), $R_{n}^{\theta }(i)$ and $R_{n}^{\phi }(i)$ are designed to be negative to $i$ and positive to $d_{{\bar{\delta }}}^{(i-1)}$. Thus, the focusing angles which are far away from the actual DoAs can be removed continuously, thereby improving the performance and reducing computational complexity.}
\item{\textbf{About} $[b-\cos\hat\theta_n^{(i-1)}]$ \textbf{and} $[b-\sin\hat\theta_n^{(i-1)}]$\textbf{:} When the absolute values of the partial derivatives of the steering vector $\mathbf{a}({{f}_{z}},{{\theta }_{n}},{{\phi }_{n}})$ in (3) to $\theta_n$ and $\phi_n$ are small, the adjacent angles are easily to be confused if noise exists, which affects the DoA estimation accuracy. Therefore, the partial derivatives of the steering vectors are also considered in the design of $R_{n}^{\theta }(i)$ and $R_{n}^{\phi }(i)$. The partial derivatives $\boldsymbol{\rho}_{z}^{n}(\tau )=\partial \mathbf{a}({{f}_{z}},{{\theta }_{n}},{{\phi }_{n}})/\partial \tau$ with $\tau\in\{\theta_n,\phi_n\}$ can be written as:
\begin{equation}
\begin{aligned}
\boldsymbol{\rho}_z^n(\theta_n)&=j\frac{2\pi rf_z}{c}\cos\theta_n\times\operatorname{diag}\bigg[\cos\left(\frac{2\pi\times 0}{M}-\phi_n\right), \dots, \cos\left(\frac{2\pi\times (M-1)}{M}-\phi_n\right)\bigg]\mathbf{a}(f_z,\theta_n,\phi_n)\text{,}
\end{aligned}
\end{equation}
\begin{equation}
\begin{aligned}
\boldsymbol{\rho}_z^n(\phi_n)&=j\frac{2\pi rf_z}{c}\sin\theta_n\times\operatorname{diag}\bigg[\sin\left(\frac{2\pi\times 0}{M}-\phi_n\right), \dots, \sin\left(\frac{2\pi\times (M-1)}{M}-\phi_n\right)\bigg]\mathbf{a}(f_z,\theta_n,\phi_n)\text{.}
\end{aligned}
\end{equation}
Therefore, $R_{n}^{\theta }(i)$ and $R_{n}^{\phi }(i)$ are respectively designed to be negatively related with $\cos\hat\theta_n^{(i-1)}$ and $\sin\hat\theta_n^{(i-1)}$ and thus $\boldsymbol{\rho}_{z}^{n}(\hat\theta_n^{(i-1)})$ and $\boldsymbol{\rho}_z^n(\hat\phi_n^{(i-1)})$, so that the radii are adaptively adjusted with $\theta_n^{(i-1)}$ and $\phi_n^{(i-1)}$ to improve the probability of the range constrained by the robustness intervals that cover the actual DoAs. Besides, $b$ in (11) and (12) is used to control the sensitivity of the radii to the partial derivatives. By adjusting the value of $b$, the degree of the influence of $\cos\hat\theta_n^{(i-1)}$ and $\sin\hat\theta_n^{(i-1)}$ on $R_{n}^{\theta }(i)$ and $R_{n}^{\phi }(i)$ is respectively changed, thereby affecting the sensitivity. With the consideration of both the sensitivity and the range corresponding to the robustness intervals that covers the actual DoAs with high probability, $b>2$ is recommended.}
\end{itemize}

Based on the aforementioned centers and radii, the robustness intervals are defined as:
\begin{equation}
\mathbb{G}_{n}^{\theta }(i)=\big[ \max [ {{0}^{\circ }},\hat{\theta }_{n}^{(i-1)}-R_{n}^{\theta }(i) ],\min [ \hat{\theta }_{n}^{(i-1)}+R_{n}^{\theta }(i),{{90}^{\circ }} ] \big]\text{,}
\end{equation}
\begin{equation}
\mathbb{G}_{n}^{\phi}(i)=\big[ \hat{\phi }_{n}^{(i-1)}-R_{n}^{\phi }(i),\hat{\phi }_{n}^{(i-1)}+R_{n}^{\phi }(i) \big]\text{,}
\end{equation}
where $\mathbb{G}_{n}^{\theta}(i)$ and $\mathbb{G}_{n}^{\phi}(i)$ respectively denote the $n$-th robustness interval of elevation and azimuth. Then, the focusing angles is obtained by sampling the range constrained by $\mathbb{G}_{n}^{\theta}(i)$ and $\mathbb{G}_{n}^{\phi}(i)$, $n=1,2,\dots,\hat{N}_{i-1}$, with the elevation step $v_\theta$ and the azimuth step $v_\phi$. Define the set of the focusing angles in the $i$-th iteration as $\mathbb{S}_\text{p}^i$. When employing the RSS method, the focusing matrices at the frequency points in $\mathbb{F}_{\text{in}}^{(i)}$ are calculated by (5) where the target matrix of SVD is replaced by $\mathbf{A}({{f}_{z}},\mathbb{S}_\text{p}^i){{\mathbf{A}}^{\text{H}}}({{f}_{0}},\mathbb{S}_\text{p}^i)$. Then, $\mathbf{R_Y}$ is computed as:
\begin{equation}
\mathbf{R_Y}=\frac{1}{K_fZ^2}\sum\limits_{f_z\in \mathbb{F}_{\text{in}}^{(i)}}[\mathbf{B}(f_z)\mathbf{X}_f(f_z)][\mathbf{B}(f_z)\mathbf{X}_f(f_z)]^\text{H}\text{.}
\end{equation}

3) \textit{MUSIC algorithm with the range-shrunk spatial spectrum}

Since the pre-estimated DoAs as well as $(\hat\theta_n^{(i-1)},\hat\phi_n^{(i-1)})$ are generally not far away from the actual DoAs \cite{ref18,ref19,ref20,ref21,ref22,ref23}, where $n=1,2,\dots,\hat{N}_{i-1}$, it is unnecessary to generate the whole spatial spectrum of the MUSIC algorithm like C-CSM. To reduce the computational burden, the range of spatial spectrum is shrunk to the regions that are constrained by $\mathbb{G}_{n}^{\theta}(i)$ and $\mathbb{G}_{n}^{\phi}(i)$, $n=1,2,\dots,\hat{N}_{i-1}$, since these regions are designed to cover the actual DoAs in most cases as mentioned in the previous part. Afterwards, the MUSIC algorithm with the range-shrunk spatial spectrum is utilized to estimate the DoAs based on $\mathbf{R_Y}$ in (17), where the spatial spectrum is only required to be computed at the sampled directions in $\mathbb{S}_\text{p}^i$.

4) \textit{Iteration termination judgement}

In each iteration, the estimation results are sent to the iteration termination judgement part. The iteration terminates when the estimation results are equal to the ones in the previous iteration. In addition, the iteration also terminates if the number of iterations reaches the upper limit number $I$, so as to avoid heavy computational burden. The results of the last iteration are output as the eventual estimated DoAs.

To sum up, the proposed method improves the robustness of C-CSM in terms of the focusing process by generating the focusing angles from the robustness intervals as mentioned in Part 2). Additionally, the computational complexity is reduced by considering less frequency points and shrinking the range of the spatial spectrum, which are respectively mentioned in Part 1) and Part 3). The procedures of RIPF-CSM with the RSS method are summarized in Algorithm 1.

\begin{algorithm}[!htb]
\renewcommand{\algorithmicrequire}{\textbf{Input:}}
\renewcommand{\algorithmicensure}{\textbf{Output:}}
\caption{RIPF-CSM with the RSS method.}
\vspace{1mm}
\begin{algorithmic}[1]
\REQUIRE{$M$, $r$, $b$, $f_0$, $Z$, $I$, $\overline{\theta_\text{e}}$, $\overline{\phi_\text{e}}$, $v_\theta$, $v_\phi$, $\mathbf{X}_f(f_1), \mathbf{X}_f(f_2), \dots, \mathbf{X}_f(f_Z)$, $f_1, f_2, \dots, f_Z$, $(\hat{\theta}_1^{(0)},\hat{\phi}_1^{(0)}), (\hat{\theta}_2^{(0)},\hat{\phi}_2^{(0)}), \dots, (\hat{\theta}_{\hat{N}_0}^{(0)},\hat{\phi}_{\hat{N}_0}^{(0)})$.}
\ENSURE{Estimated DoAs of the wideband signal sources}.
\STATE \textbf{\textit{Initialization}:} Compute $d_\theta$ and $d_\phi$ in (9), and set $i$ as $1$.
\STATE \textbf{while} $i \le I$ \textbf{do} \\
\hspace{0.22cm} \textbf{\textit{Frequency point selection}:}
\STATE \hspace{0.22cm} \textbf{if} $i=1$ \textbf{then}
\STATE \hspace{0.59cm} Select a frequency point randomly and form $\mathbb{F}_\text{in}^{(1)}$.
\STATE \hspace{0.22cm} \textbf{else}
\STATE \hspace{0.59cm} Calculate $\varDelta_f^{(i)}$ by (9).
\STATE \hspace{0.59cm} Randomly select $\varDelta_f^{(i)}$ frequency points which are not in $\mathbb{F}_\text{in}^{(i-1)}$, and then merge them with $\mathbb{F}_\text{in}^{(i-1)}$ to form $\mathbb{F}_\text{in}^{(i)}$.
\STATE \hspace{0.22cm} \textbf{end if} \\
\hspace{0.22cm} \textbf{\textit{Focusing process}:}
\STATE \hspace{0.22cm} Calculate $R_n^\theta(i)$ and $R_n^\phi(i)$ by (11) and (12), then generate $\mathbb{G}_n^\theta(i)$ and $\mathbb{G}_n^\phi(i)$ by (15) and (16).
\STATE \hspace{0.22cm} Form $\mathbb{S}_\text{p}^i$ by sampling $\mathbb{G}_n^\theta(i)$ and $\mathbb{G}_n^\phi(i)$ with $v_\theta$ and $v_\phi$.
\STATE \hspace{0.22cm} Compute{\hfill}$\mathbf{A}(f_z,\mathbb{S}_\text{p}^i)\mathbf{A}^\text{H}(f_0,\mathbb{S}_\text{p}^i)${\hfill}at{\hfill}all{\hfill}$f_z\in\mathbb{F}_\text{in}^{(i)}$,{\hfill}and{\hfill}then{\hfill}calculate{\hfill}the{\hfill}corresponding $\mathbf{B}(f_z)${\hfill}by{\hfill}substituting{\hfill}the{\hfill}SVD \\ \hspace{0.22cm} results of $\mathbf{A}(f_z,\mathbb{S}_\text{p}^i)\mathbf{A}^\text{H}(f_0,\mathbb{S}_\text{p}^i)$ into (5).
\STATE \hspace{0.22cm} Compute $\mathbf{R_Y}$ by (17). \\
\hspace{0.22cm} \textbf{\textit{MUSIC algorithm}:}
\STATE \hspace{0.22cm} Perform EVD on $\mathbf{R_Y}$, and identify $\hat{N}_i$ in (8) by the difference between the eigenvalues.
\STATE \hspace{0.22cm} Form $\mathbf{E}_\text{n}$ by the eigenvectors of the noise-subspace.
\STATE \hspace{0.22cm} Sample{\hfill}the{\hfill}range{\hfill}constrained{\hfill}by{\hfill}$\mathbb{G}_{n}^{\theta}(i)${\hfill}and{\hfill}$\mathbb{G}_{n}^{\phi}(i)$,{\hfill}$n=1,2,\dots,\hat{N}_{i-1}$,{\hfill}with{\hfill}$v_\theta${\hfill}and{\hfill}$v_\phi$,{\hfill}and{\hfill}then{\hfill}compute{\hfill}the{\hfill}spatial \\
\hspace{0.22cm} spectrum at the sampled directions by (7).
\STATE \hspace{0.22cm} Estimate $(\hat\theta_1^{(i)},\hat\phi_1^{(i)}),(\hat\theta_2^{(i)},\hat\phi_2^{(i)}),\dots,(\hat\theta_{\hat{N}_i}^{(i)},\hat\phi_{\hat{N}_i}^{(i)})$ by searching for the largest $\hat{N}_i$ peaks in the spatial spectrum. \\
\hspace{0.22cm} \textbf{\textit{Iteration termination judgement}:}
\STATE \hspace{0.22cm} Calculate $\bar{\delta}^{(i)}$ by (8).
\STATE \hspace{0.22cm} \textbf{if} $\bar{\delta}^{(i)}=0$ \textbf{and} $\hat{N}_{i-1}=\hat{N}_i$ \textbf{then}
\STATE \hspace{0.59cm} {\bf{exit while}}
\STATE \hspace{0.22cm} \textbf{end if}
\STATE \hspace{0.22cm} Compute $d_{\bar{\delta}}^{(i)}$ in (9), and update $i$ by $i+1$.
\STATE \textbf{end while}
\RETURN $(\hat\theta_1^{(i)},\hat\phi_1^{(i)}),(\hat\theta_2^{(i)},\hat\phi_2^{(i)}),\dots,(\hat\theta_{\hat{N}_i}^{(i)},\hat\phi_{\hat{N}_i}^{(i)})$.
\end{algorithmic}
\label{alg1}
\end{algorithm}

\subsection{Computational Complexity}

In this subsection, the computational complexities of the proposed RIPF-CSM and some benchmark methods are discussed, where the number of floating-point operations (FLOPs) is used to represent the computational complexity.

For clarity, the computational complexity of RIPF-CSM with the RSS method in a single iteration is analysed first. The complexities of some special operations including the trigonometric function, SVD and EVD in Algorithm 1 are discussed as follows. The trigonometric function requires a small and constant amount of FLOPs with a certain demand of precision \cite{ref32}. Besides, the computational complexity of SVD is generally twice as much as that of EVD \cite{ref33} whose complexity is expressed as $\mathcal{O}(M^3)$ FLOPs \cite{ref34}. Considering these operations together with the remaining arithmetic operations and logic judgements, the overall computational complexity of the RIPF-CSM with the RSS method is $\mathcal{O}\Big(2Z_{\text{in}}^{(i)}{{M}^{2}}[M+8{{K}_{f}}+16\sum\limits_{n=1}^{\hat{N}_{i-1}}R_n^\theta(i)R_n^\phi(i)/(v_\theta v_\phi)]\Big)$ by summarizing the complexities of all operations and just keeping the highest order terms of the parameters in the expression.

To better reveal the complexity advantage of the proposed method, the computational complexities of some benchmark methods are also counted, which include C-CSM, the C-CSM with specific extra focusing angles (SE-CSM) \cite{ref19}, R-CSM \cite{ref22} and I-2D-CSM \cite{ref23}. For fair comparison, all of their acquisition processes of the pre-estimated DoAs are not taken into account in this paper, and the original versions of these methods for 1D DoA estimation are extended to the corresponding versions for 2D DoA estimation as shown in Appendix A. In addition, the focusing matrices of the benchmark methods are generated by the RSS method. There are three differences in terms of complexity between RIPF-CSM and the other methods. First, the benchmark methods use all of the $Z$ frequency points rather than partial frequency points. Second, there exists differences in the number of focusing angles between RIPF-CSM and the benchmark methods, which are controlled by the radii of robustness intervals. The radii of the robustness intervals of C-CSM and SE-CSM are respectively equivalent to $0$ and $1$. Additionally, the elevation and azimuth robustness interval radii of R-CSM are respectively denoted as $R{{_{n}^{\theta }}^{\prime }}(i)$ and $R{{_{n}^{\phi }}^{\prime }}(i)$, and those of I-2D-CSM are respectively denoted as $R{{_{n}^{\theta }}^{\prime\prime }}(i)$ and $R{{_{n}^{\phi }}^{\prime\prime }}(i)$, whose expressions are given in Appendix A. Third, the range of the spatial spectrum in the benchmark methods is constrained by $\theta\in [0^\circ,90^\circ]$ and $\phi\in[0^\circ,360^\circ)$ rather than the proposed robustness intervals. The numbers of elevation and azimuth angles sampled within the aforementioned range are respectively denoted as ${{L}_{\theta }}={{90}^{\circ }}/{{v}_{\theta }}+1$ and $L_\phi={{360}^{\circ }}/{{v}_{\phi }}$. According to the above discussions, the computational complexity of each method in a single iteration is summarized in Table 1.

\begin{table}[!tb]
\caption{Computational complexities of different CSM methods in a single iteration}
\centering
\small
\renewcommand\arraystretch{1}
\begin{tabular}{|m{2cm}<{\centering}|m{10cm}<{\centering}|}
\hline
\textbf{Method} & \textbf{Computational complexity in a single iteration}\\
\hline
C-CSM & \vspace{0.5mm}$\mathcal{O}\Big(2\bm{Z}M^2(M+8K_f+\bm{4\hat{N}_{i-1}})+\bm{8M^2L_{\theta}L_{\phi}}\Big)$ \\[0.5mm]
\hline
SE-CSM & \vspace{0.5mm}$\mathcal{O}\Big(2\bm{Z}M^2(M+8K_f+\bm{20\hat{N}_{i-1}})+\bm{8M^2L_{\theta}L_{\phi}}\Big)$ \\[0.5mm]
\hline
R-CSM & \vspace{0.5mm}$\mathcal{O}\Big(2\bm{Z}M^2[M+8K_f+\bm{16\sum\limits_{n=1}^{\hat{N}_{i-1}}{R_n^{\theta}}^{\prime}(i){R_n^{\phi}}^{\prime}(i)/(v_{\theta}v_{\phi})}]+\bm{8M^2L_{\theta}L_{\phi}}\Big)$ \\[1.5mm]
\hline
I-2D-CSM & \vspace{0.5mm}$\mathcal{O}\Big(2\bm{Z}M^2[M+8K_f+\bm{16\sum\limits_{n=1}^{\hat{N}_{i-1}}{R_n^{\theta}}^{\prime\prime}(i){R_n^{\phi}}^{\prime\prime}(i)/(v_{\theta}v_{\phi})}]+\bm{8M^2L_{\theta}L_{\phi}}\Big)$ \\[1.5mm]
\hline
RIPF-CSM & \vspace{0.5mm}$\mathcal{O}\Big(2\bm{Z_\textbf{in}^{(i)}}M^2[M+8K_f+\bm{16\sum\limits_{n=1}^{\hat{N}_{i-1}}R_n^{\theta}(i)R_n^{\phi}(i)/(v_{\theta}v_{\phi})}]\Big)$ \\[1.5mm]
\hline
\end{tabular}
\vspace{-2mm}
\end{table}

It is observed from Table 1 that there is only one different term in the complexity expressions of the benchmark methods. Since $v_\theta$ and $v_\phi$ are usually not greater than $1^\circ$ \cite{ref4,ref11,ref12} and R-CSM requires about less than five iterations to converge \cite{ref22}, $R{{_{n}^{\theta }}^{\prime }}(i)R{{_{n}^{\phi }}^{\prime }}(i)/({{v}_{\theta }}{{v}_{\phi }})>1/4$ easily holds. In this sense, The computational complexity of C-CSM is lower than that of R-CSM. According to Appendix A and $v_\theta,v_\phi\le 1^{\circ}$, $i_\text{s}^\theta,i_\text{s}^\phi\le 2$ can be derived, then ${R_{n}^{\theta }}^{\prime \prime }(i){R_{n}^{\phi }}^{\prime \prime }(i)/({{v}_{\theta }}{{v}_{\phi }})>1/4$ holds. Thus, the computational burden of I-2D-CSM is heavier than that of C-CSM. Considering the complexity of SE-CSM which is obviously higher than that of C-CSM according to Table 1, the computational complexity of C-CSM in a single iteration is the lowest among the benchmark methods. To guarantee the complexity advantage of RIPF-CSM in a single iteration when compared with C-CSM, a parameter constraint is derived, i.e., $\big[4I^2+\ln 100(d_\theta+d_\phi)[(d_\theta+d_\phi)I+2Z]\big]\times\big[M+8K_f+(2\ln 100)^2N\overline{\theta_\text{e}}\hspace{0.4mm}\overline{\phi_\text{e}}(d_\theta+d_\phi)^2b(b-1)/(v_\theta v_\phi I^2)\big]<4I^2[Z(M+4N+8K_f)+4L_\theta L_\phi]$, which can be easily satisfied and is discussed in Appendix B. Following this parameter constraint, the computational complexity of RIPF-CSM in a single iteration can be lower than that of any benchmark method.

Verified by the simulations with the general parameters of practical scenarios, the number of iterations of RIPF-CSM is significantly less than that of C-CSM and SE-CSM, and approximately equal to that of R-CSM and I-2D-CSM. Thus, with the consideration of the computational complexity in a single iteration, RIPF-CSM enjoys lower computational burden when compared with the benchmarks.

\section{Cramér-Rao Bound}
This section first derives the CRB of the DoA estimation under the signal model in this paper. Then, the guidelines for practical application are discussed based on the derived CRB.

\subsection{Derivation of CRB}
In the derivation of CRB, the first and the most important step is to derive the Fisher information matrix $\mathbf{F}$ \cite{ref9}. The CRBs of $\theta_1,\theta_2,\dots,\theta_N,\phi_1,\phi_2,\dots,\phi_N$ are respectively the main diagonal elements of the $2N$-order square matrix which locates in the lower right corner of the inverse of $\mathbf{F}$. Thus, the derivation of $\mathbf{F}$ is first described.

Denote the vector of the $k$-th snapshot in frequency domain of ${{\mathbf{X}}_{f}}({{f}_{z}})$, ${{\mathbf{S}}_{f}}({{f}_{z}})$ and ${{\mathbf{W}}_{f}}({{f}_{z}})$ as ${{\mathbf{x}}_{f}}(k,{{f}_{z}})\in\mathbb{C}^{M\times 1}$, ${{\mathbf{s}}_{f}}(k,{{f}_{z}})\in\mathbb{C}^{N\times 1}$ and $\mathbf{w}(k,{{f}_{z}})\in\mathbb{C}^{M\times 1}$ respectively, where $k\in\{1,2,\dots,K_f\}$ and $z\in \{1,2,\dots,Z\}$, where $\mathbf{s}_f(k,f_z)=[S(1,k,f_z),S(2,k,f_z),\dots,S(N,k,f_z)]^{\text{T}}$, and $S(n,k,{{f}_{z}})$ represents the $k$-th frequency domain snapshot of the source signal of the $n$-th DoA at the frequency point $f_z$, $n\in\{1,2,\dots,N\}$. According to (2), these vectors satisfy the following equation:
\begin{equation}
{{\mathbf{x}}_{f}}(k,{{f}_{z}})=\mathbf{A}({{f}_{z}},\mathbb{S}){{\mathbf{s}}_{f}}(k,{{f}_{z}})+\mathbf{w}(k,{{f}_{z}})\text{.}
\end{equation}
For arbitrary frequency point and snapshot in frequency domain, $\mathbf{w}(k,{{f}_{z}})$ obeys $M$-dimensional complex Gaussian distribution, whose covariance matrix is $\mathbf{\Sigma }=\mathcal{E}[\mathbf{w}(k,f_z)\mathbf{w}^\text{H}(k,f_z)]=Z\sigma^2\mathbf{I}_M$. In addition, denote the probability density function of $\mathbf{w}(k,{{f}_{z}})$ as $p(k,{{f}_{z}})$, and then define the natural logarithm of the joint probability density function $P$ of $\mathbf{w}(k,{{f}_{z}})$ with $k=1,2,\dots,K_f$ and $z=1,2,\dots,Z$ as:
\begin{equation}
\ln P=\ln \prod\limits_{k=1}^{{{K}_{f}}}{\prod\limits_{z=1}^{Z}{p(k,{{f}_{z}})}}=-{{K}_{f}}MZ( \ln \pi Z+\ln {{\sigma }^{2}})-\frac{1}{Z{{\sigma }^{2}}}\sum\limits_{k=1}^{{{K}_{f}}}{\sum\limits_{z=1}^{Z}{{{\mathbf{w}}^{\text{H}}}( k,{{f}_{z}})\mathbf{w}( k,{{f}_{z}})}}\text{.}
\end{equation}
According to (3), (18) and (19), the unknown parameters in $\ln P$ include $\sigma^2$, $\operatorname{Re}[S(n,k,{{f}_{z}})]$, $\operatorname{Im}[S(n,k,{{f}_{z}})]$, $\theta_n$ and $\phi_n$, where $n=1,2,\dots,N$, $k=1,2,\dots,K_f$ and $z=1,2,\dots,Z$. According to Appendix C, with the definitions that $\mathbf{\hat{s}}_f(k,f_z)=\operatorname{Re}[\mathbf{s}_f(k,f_z)]$, $\mathbf{\check{s}}_f(k,f_z)=\operatorname{Im}[\mathbf{s}_f(k,f_z)]$, $\mathbf{\bar{s}}(f_z)=[\mathbf{\hat{s}}_f^\text{T}(1,f_z),\mathbf{\check{s}}_f^\text{T}(1,f_z),\dots,\mathbf{\hat{s}}_f^\text{T}(K_f,f_z),\mathbf{\check{s}}_f^\text{T}(K_f,f_z)]$, $\boldsymbol{\zeta}=[\theta_1,\dots,\theta_N,\phi_1,\dots,\phi_N]$, and $\boldsymbol{\psi}^{\text{T}}=\partial \ln P/\partial [\sigma^2,\mathbf{\bar{s}}(f_1),\mathbf{\bar{s}}(f_2),\dots,\mathbf{\bar{s}}(f_Z),\boldsymbol{\zeta }]$, the Fisher information matrix $\mathbf{F}=\mathcal{E}[\boldsymbol{\psi}{{\boldsymbol{\psi}}^{\text{T}}}]$ \cite{ref9} is derived as:
\begin{equation}
\mathbf{F}=\left[
\setlength{\arraycolsep}{3pt}
\renewcommand\arraystretch{0.9}
\begin{array}{c:cccc}
\frac{{{K}_{f}}MZ}{{{\sigma }^{4}}} & {} & {} & 0 & {}  \\
\hdashline
{} & \mathbf{\Lambda }_{\text{F}}^{d}( {{f}_{1}}) & {} & {} & {{\mathbf{\Phi }}_{\text{F}}}( {{f}_{1}})  \\
0 & {} & \ddots  & {} & \vdots   \\
{} & {} & {} & \mathbf{\Lambda }_{\text{F}}^{d}( {{f}_{Z}}) & {{\mathbf{\Phi }}_{\text{F}}}( {{f}_{Z}})  \\
{} & \mathbf{\Phi }_\text{F}^\text{T}( {{f}_{1}}) & \cdots  & \mathbf{\Phi }_\text{F}^\text{T}({{f}_{Z}}) & \mathbf{\Gamma }
\end{array}
\right]\text{,}
\end{equation}
where
\begin{equation}
\mathbf{\Lambda }( {{f}_{z}} )=\frac{2}{Z{{\sigma }^{2}}}{{\mathbf{A}}^{\text{H}}}( {{f}_{z}} )\mathbf{A}( {{f}_{z}})\text{,}
\end{equation}
\begin{equation}
{{\mathbf{\Lambda }}_{\text{F}}}( {{f}_{z}} )=
\renewcommand\arraystretch{0.9}
\begin{bmatrix}
\operatorname{Re}\left[ \mathbf{\Lambda }( {{f}_{z}}) \right] & -\operatorname{Im}\left[ \mathbf{\Lambda }( {{f}_{z}}) \right]  \\
\operatorname{Im}\left[ \mathbf{\Lambda }( {{f}_{z}}) \right] & \operatorname{Re}\left[ \mathbf{\Lambda }( {{f}_{z}}) \right]  \\
\end{bmatrix}\text{,}
\end{equation}
\begin{equation}
\mathbf{\Lambda }_{\text{F}}^{d}({{f}_{z}})=\operatorname{diag}[\overbrace{{{\mathbf{\Lambda }}_{\text{F}}}({{f}_{z}}),\dots,{{\mathbf{\Lambda }}_{\text{F}}}({{f}_{z}})}^{{K}_{f}} ]\text{,}
\end{equation}
\begin{equation}
\mathbf{\Xi }(k,{{f}_{z}})=\operatorname{diag}[S( 1,k,{{f}_{z}} ),\dots,S( N,k,{{f}_{z}}),S( 1,k,{{f}_{z}}),\dots,S( N,k,{{f}_{z}})]\text{,}
\end{equation}
\begin{equation}
\mathbf{D}\left( {{f}_{z}} \right)=[\boldsymbol{\rho }_{z}^{n}({{\theta }_{1}}),\dots,\boldsymbol{\rho }_{z}^{n}({{\theta }_{N}}),\boldsymbol{\rho }_{z}^{n}({{\phi }_{1}}),\dots,\boldsymbol{\rho }_{z}^{n}({{\phi }_{N}})]\text{,}
\end{equation}
\begin{equation}
\mathbf{\Phi }( k,{{f}_{z}})=\frac{2}{Z{{\sigma }^{2}}}{{\mathbf{A}}^{\text{H}}}( {{f}_{z}})\mathbf{D}( {{f}_{z}})\mathbf{\Xi }( k,{{f}_{z}})\text{,}
\end{equation}
\begin{equation}
{{\mathbf{\Phi }}_{\text{F}}}( {{f}_{z}})=\big[\operatorname{Re}\left[ {{\mathbf{\Phi }}^{\text{T}}}( 1,{{f}_{z}}) \right],\operatorname{Im}\left[ {{\mathbf{\Phi }}^{\text{T}}}( 1,{{f}_{z}}) \right],\dots,\operatorname{Re}\left[ {{\mathbf{\Phi }}^{\text{T}}}( {{K}_{f}},{{f}_{z}}) \right],\operatorname{Im}\left[ {{\mathbf{\Phi }}^{\text{T}}}( {{K}_{f}},{{f}_{z}}) \right] \big]^\text{T}\text{,}
\end{equation}
\begin{equation}
\mathbf{\Gamma }=\mathcal{E}\left[ \mathbf{d}_{\boldsymbol{\zeta}}\mathbf{d}_{\boldsymbol{\zeta}}^{\text{T}} \right]\text{,}
\end{equation}
and $\mathbf{d}_{\boldsymbol{\zeta}}=\partial\ln P/\partial\boldsymbol{\zeta}$. The $2N$-order square matrix in the lower right corner of $\mathbf{F}^{-1}$ is denoted as ${{\mathbf{F}}_{\text{inv}}}( \boldsymbol{\zeta })$, which can be calculated by the inverse of block matrix as:
\vspace{0mm}
\begin{equation}
{{\mathbf{F}}_{\text{inv}}}(\boldsymbol{\zeta })=\frac{Z\sigma^2}{2}\Big[ \sum\limits_{k=1}^{K_f}\sum\limits_{z=1}^{Z}\operatorname{Re}\big[ \mathbf{\Xi}^\text{H}(k,f_z)\mathbf{D}^\text{H}(f_z)\times\mathbf{P}_\mathbf{A}^{\bot}(f_z)\mathbf{D}(f_z)\mathbf{\Xi}(k,f_z)\big] \Big]^{-1}\text{,}
\end{equation}
where $\mathbf{P}_\mathbf{A}^\bot(f_z)=\mathbf{I}_M-\mathbf{P_A}(f_z)=\mathbf{I}_M-\mathbf{A}(f_z)[\mathbf{A}^\text{H}(f_z)\mathbf{A}(f_z)]^{-1}\mathbf{A}^\text{H}(f_z)$, and $\mathbf{P_A}(f_z)$ is an orthogonal projection matrix. As a result, the CRBs of ${{\theta }_{1}},\dots,{{\theta }_{N}},{{\phi }_{1}},\dots,{{\phi }_{N}}$ are the elements in the main diagonal of ${{\mathbf{F}}_{\text{inv}}}( \boldsymbol{\zeta })$.

\subsection{Guidelines}
To obtain the DoA estimation with higher accuracy, a smaller CRB is more favorable. With the expectation of improving the estimation performance, some guidelines to decrease the CRB are discussed as follows.
\begin{itemize}
\item{\textbf{The sampling duration:} As shown in (29), the larger number $K_f$ of the frequency domain snapshots is, the smaller CRBs of DoAs are. Therefore, the DoA estimation performance can be improved by increasing the sampling duration $t_0$ when the sampling rate $f_\text{S}$ remains unchanged.}
\item{\textbf{The signal bandwidth:} According to the Nyquist sampling theorem, wider signal bandwidth requires higher sampling rate, which implies an increased $K_f$ when $t_0$ remains unchanged, thereby decreasing the CRB. Thus, DoA estimation can benefit from the wider bandwidth.}
\item{\textbf{The signal power:} (29) can be rewritten as:
\begin{equation}
\mathbf{F}_\text{inv}(\boldsymbol\zeta)=\frac{Z\sigma^2}{2}\bigg[ \sum\limits_{k=1}^{K_f}\sum\limits_{z=1}^{Z}\operatorname{Re}\big[\big( \mathbf{D}^{\text{H}}(f_z)\mathbf{P}_\mathbf{A}^{\bot}(f_z)\times\mathbf{D}(f_z) \big) \odot \mathbf{R}_{2\mathbf{S}}(k,f_z) \big] \bigg]^{-1}\text{,}
\end{equation}
where ${{\mathbf{R}}_{2\mathbf{S}}}( k,{{f}_{z}})={{\mathbf{\Xi }}^{\text{H}}}(k,{{f}_{z}}) \mathbf{1}_{2N} \mathbf{\Xi }( k,{{f}_{z}})$. It is easy to know that the increase of the signal power enlarges the magnitudes of the elements in ${\mathbf{R}}_{2\mathbf{S}}(k,{{f}_{z}})$, and thus the CRB decreases. Therefore, high signal power helps improve the DoA estimation performance.}
\item{\textbf{The number of array elements:} ${{\mathbf{D}}^{\text{H}}}( {{f}_{z}} )\mathbf{P}_{\mathbf{A}}^{\bot }( {{f}_{z}} )\mathbf{D}( {{f}_{z}} )$ in (29) can be expanded as:
\begin{equation}
{{\mathbf{D}}^{\text{H}}}( {{f}_{z}} )\mathbf{P}_{\mathbf{A}}^{\bot }( {{f}_{z}} )\mathbf{D}( {{f}_{z}} )={{\mathbf{D}}^{\text{H}}}( {{f}_{z}} )\mathbf{D}( {{f}_{z}} )-{{\mathbf{D}}^{\text{H}}}( {{f}_{z}}){{\mathbf{P}}_{\mathbf{A}}}( {{f}_{z}})\mathbf{D}( {{f}_{z}} )\text{,}
\end{equation}
where the element in the $q_1$-th row and $q_2$-th column of ${{\mathbf{D}}^{\text{H}}}\left( {{f}_{z}} \right)\mathbf{D}\left( {{f}_{z}} \right)$ can be written as:
\begin{equation}
\left[ {{\mathbf{D}}^{\text{H}}}( {{f}_{z}})\mathbf{D}( {{f}_{z}}) \right]_{q_1q_2}
=
\begin{cases}
(2\pi rf_z/c)^2\operatorname{Pe}_1(r,f_z,\theta_{q_1},\theta_{q_2}),&\text{if}~{{q}_{1}}\le N,{{q}_{2}}\le N\text{,}\\
(2\pi rf_z/c)^2\operatorname{Pe}_2(r,f_z,\theta_{q_1},\phi_{q_2-N}),&\text{if}~{{q}_{1}}\le N,{{q}_{2}}> N\text{,}\\
(2\pi rf_z/c)^2\operatorname{Pe}_3(r,f_z,\phi_{q_1-N},\theta_{q_2}),&\text{if}~{{q}_{1}}> N,{{q}_{2}}\le N\text{,}\\
(2\pi rf_z/c)^2\operatorname{Pe}_4(r,f_z,\phi_{q_1-N},\phi_{q_2-N}),&\text{otherwise,}
\end{cases}
\end{equation}
where $q_1,q_2\in\{1,\dots,2N\}$ and $\operatorname{Pe}_q(r,{{f}_{z}},{\tau }_{1},{\tau }_{2})$ represents a periodic function which is the product of multiple periodic functions according to (13) and (14), $\tau_1,\tau_2\in\{\theta_1,\dots,\theta_N,\phi_1,\dots,\phi_N\}$, $q\in\{1,2,3,4\}$, and its period is related with $r$, $f_z$, $\tau_1$, $\tau_2$. Due to the periodicity of $\operatorname{Pe}_q(r,f_z,\tau_1,\tau_2)$, its effect on the magnitude of $[{{\mathbf{D}}^{\text{H}}}({{f}_{z}})\mathbf{D}({{f}_{z}})]_{q_1q_2}$ is ignored, and thus its detailed expressions are omitted for conciseness. Considering the coefficient $(2\pi rf_z/c)^2$ in (32), the magnitude of each element in ${{\mathbf{D}}^{\text{H}}}( {{f}_{z}} )\mathbf{D}( {{f}_{z}} )$ is positively related with the radius $r$ of UCA which grows with the number $M$ of array elements. Since $\mathbf{P_A}(f_z)$ is an idempotent Hermitian matrix, ${{\mathbf{D}}^{\text{H}}}( {{f}_{z}} )\mathbf{P_A}(f_z)\mathbf{D}( {{f}_{z}} )$ can be expanded as:
\begin{equation}
{{\mathbf{D}}^{\text{H}}}( {{f}_{z}} )\mathbf{P}_{\mathbf{A}}^{\bot }( {{f}_{z}} )\mathbf{D}( {{f}_{z}} )={{\mathbf{D}}^{\text{H}}}( {{f}_{z}} )\mathbf{D}( {{f}_{z}} )-{{\mathbf{D}}^{\text{H}}}( {{f}_{z}}){{\mathbf{P}}_{\mathbf{A}}}( {{f}_{z}})\mathbf{D}( {{f}_{z}} )\text{,}
\end{equation}
where ${{\mathbf{P}}_{\mathbf{A }}}({{f}_{z}})\mathbf{D}({{f}_{z}})$ is equivalent to projecting $\mathbf{D}({{f}_{z}})$ into a low-dimensional subspace. Accordingly, the result of (33) is the product of the coefficient $(2\pi rf_z/c)^2$ and a certain periodic function like (32). Thus, the CRB is negatively related with the number $M$ of array elements, so that the DoA estimation performance can take advantage of more array elements.}
\end{itemize}

According to the above discussions, the CRB is generally inversely proportional to $K_f$, $r$ and thus $1/\sin(\pi/M)$, and the magnitudes of the elements in ${{\mathbf{R}}_{2\mathbf{S}}}( k,{{f}_{z}})$. In addition, the derivatives of $1/{{K}_{f}}$, $\sin(\pi/M)$ and $1/P_S$ with respect to their variables are $-1/{K}_{f}^2$, $-\cos(\pi/M)/M^2$ and $-1/P_S^2$ respectively, where $P_S$ represents the average power of the signals. Therefore, by adopting the strategy that increases the corresponding variables of these derivatives especially the smallest derivative, the accuracy of the wideband 2D DoA estimation with a UCA can be improved effectively.

\section{Simulation Results}
In this section, the numerical simulation results of the proposed RIPF-CSM method are provided. Its DoA estimation performance and computational complexity are compared with the aforementioned benchmark methods, i.e., C-CSM with a single iteration (C-CSM ($i=1$)), C-CSM, SE-CSM, R-CSM and I-2D-CSM.

In the simulations, root mean square error (RMSE) and successful detection probability (SDP) \cite{ref35,ref36,ref37} are used as the metrics for the DoA estimation performance. Specifically, RMSE indicates the degree of the deviation between the estimation results and the actual DoAs, and it is defined as:
\begin{equation}
\text{RMSE}=\sqrt{\frac{\sum\limits_{\kappa =1}^{\mathcal{M}}{\sum\limits_{n=1}^{N}{{{( {{{\hat{\theta }}}_{n}}( \kappa  )-{{\theta }_{n}})}^{2}}+{{( {{{\hat{\phi }}}_{n}}( \kappa  )-{{\phi }_{n}} )}^{2}}}}}{\mathcal{M}N}}\text{,}
\end{equation}
where $\mathcal{M}$ is the number of Monte Carlo experiments, ${{\hat{\theta }}_{n}}(\kappa)$ and ${{\hat{\phi }}_{n}}(\kappa)$ respectively denote the elevation and azimuth of the $n$-th estimated DoA in the $\kappa$-th experiment. Additionally, the RMSE corresponding to the average CRB of multiple DoAs is calculated as:
\begin{equation}
\text{RMS}{{\text{E}}_{\text{CRB}}}=\sqrt{\frac{\sum\limits_{v=1}^{V}{\operatorname{trace}\left[ {{\mathbf{F}}_{\text{inv}}}( {{\boldsymbol{\zeta }}_{v}}) \right]}}{VN}}\text{,}
\end{equation}
where the DoAs considered simultaneously are defined as a DoA group, $V$ represents the number of DoA groups, and $\boldsymbol\zeta_v$ stands for the angle vector $\boldsymbol\zeta$ of the $v$-th DoA group. Moreover, SDP is defined as:
\begin{equation}
\text{SDP}=\mathcal{P}\left[\left\lvert {{{\hat{\theta }}}_{n}}-{{\theta }_{n}} \right\rvert+\left\lvert {{{\hat{\phi }}}_{n}}-{{\phi }_{n}} \right\rvert\le {{v}_{\theta }}+{{v}_{\phi }}\right]\text{.}
\end{equation}
which represents the probability that the DoA estimation results are close to the actual DoAs, which is used to reflect the validity of DoA estimation directly.

In addition, the running time is used to reflect the computational complexity of a method. The simulation platform is MatlabR2020a on a computer with 2.10GHz AMD Ryzen 5-4600U CPU and 16GB random access memory.

\subsection{Simulation Settings}

In the simulations, the number $M$ of UCA antenna elements is set as $5$. The speed of light $c=3\times {{10}^{8}}\text{m/s}$. The received signals are linear frequency modulated signals with ${{f}_{0}}=30\text{GHz}$, which are converted to analytic signals. The bandwidth $B$ of them and the sampling rate $f_\text{S}$ are respectively set as $9\text{GHz}$ and $11.25\text{GHz}$, and the sampling duration is ${{t}_{0}}=10\mu\text{s}$. Then, the radius of UCA is computed as $r=c/[4({{f}_{0}}+B/2)\sin(\pi/M)]$. The scenario with approximately coherent signals is considered, i.e., the same signal arrives at the antenna array from different paths successively. It is more difficult for such scenario to estimate the DoAs accurately than the scenario with uncorrelated signals, thereby revealing the universal applicability of RIPF-CSM. The time interval between the paths with adjacent arriving time are set to be $1\text{ns}$. To display the DoA estimation performance and computational complexity of each method under different number $N$ of paths, the results with $N=1,2,3$ are respectively given. In addition, for each case of $N$, three different DoA groups are respectively selected as shown in Table 2. The simulation results of different groups with the same $N$ are counted together.

Furthermore, the performance and the computational complexity are given for different signal-to-noise ratio (SNR), $\overline{\theta_\text{e}}$, $\overline{\phi_\text{e}}$, $Z$, $v_\theta$ and $v_\phi$. The default values of these parameters are $\text{SNR}=10\text{dB}$, $\overline{\theta_\text{e}}=\overline{\phi_\text{e}}={{3}^{\circ }}$, $Z=32$ and ${{v}_{\theta }}={{v}_{\phi }}={{0.2}^{\circ }}$. $Z\ge 32$ is guaranteed and thus the bandwidth corresponding to each frequency point is far less than the central frequency ${{f}_{0}}$, which can be considered as narrowband. Moreover, $b=3$ and $I=15$ are set in the simulations. Each simulation result is obtained by performing $\mathcal{M}=300$ Monte Carlo experiments.

\begin{table}[!tb]
\caption{Ground-truth DoAs in different groups}
\centering
\renewcommand\arraystretch{0.9}
\small
\begin{tabular}{|m{0.7cm}<{\centering}|m{1cm}<{\centering}|m{6cm}<{\centering}|}
\hline
$N$ & \textbf{Group} &  \textbf{DoA} \\
\hline
\multirow{3}{*}{$1$} & 1 & $(60^\circ,150^\circ)$ \\
\cline{2-3}
& 2 & $(33^\circ,50^\circ)$ \\
\cline{2-3}
& 3 & $(28^\circ,230^\circ)$ \\
\hline
\multirow{3}{*}{$2$} & 1 & $(60^\circ,150^\circ)$, $(20^\circ,45^\circ)$ \\
\cline{2-3}
& 2 & $(40^\circ,175^\circ)$, $(70^\circ,250^\circ)$ \\
\cline{2-3}
& 3 & $(25^\circ,230^\circ)$, $(65^\circ,150^\circ)$ \\
\hline
\multirow{3}{*}{$3$} & 1 & $(60^\circ,150^\circ)$, $(30^\circ,95^\circ)$, $(45^\circ,300^\circ)$ \\
\cline{2-3}
& 2 & $(30^\circ,50^\circ)$, $(40^\circ,190^\circ)$, $(70^\circ,250^\circ)$ \\
\cline{2-3}
& 3 & $(25^\circ,230^\circ)$, $(65^\circ,150^\circ)$, $(35^\circ,60^\circ)$ \\
\hline
\end{tabular}
\vspace{-2mm}
\end{table}

\subsection{RMSE and SDP Versus SNR}

This subsection gives the RMSEs and the SDPs of different methods with different SNRs and $N$ as shown in Fig. 3, where Figs. 3 (a)-(c) and Figs. 3 (d)-(f) respectively show the RMSEs and the SDPs with $\text{SNR}\in[-10\text{dB},20\text{dB}]$, Figs. 3 (a) and (d) are with $N=1$, Figs. 3 (b) and (e) are with $N=2$, Figs. 3 (c) and (f) are with $N=3$. The RMSEs of the average CRBs are also depicted as a benchmark in Figs. 3 (a)-(c).

As illustrated in Fig. 3, although the estimation accuracy of all the methods becomes worse with a larger $N$, the RMSE and the SDP of RIPF-CSM are always better than those of other methods. Moreover, RIPF-CSM achieves excellent performance which is fairly close to CRB when $\text{SNR}\in[-6\text{dB},-2\text{dB}]$ in Figs. 3 (a)-(c). Since the difference of the performance between the proposed method and the benchmark methods is mainly affected by the focusing process, it implies that the focusing process with the robustness intervals of RIPF-CSM generates more effective focusing matrices, thereby improving the DoA estimation performance.

In addition, it is observed from Fig. 3 that when SNR increases, the RMSE of the proposed method decreases, and the SDP of it increases. Fig. 3 also shows that RIPF-CSM requires lower SNR to achieve the same estimation performance as that of any benchmark method, which indicates the favorable robustness of RIPF-CSM. Furthermore, as shown in Fig. 3, the DoA estimation performance of all methods approximately converge when SNR is high. This phenomenon is caused by the characteristic of the MUSIC algorithm that utilizes the orthogonality between the signal-subspace and the noise-subspace \cite{ref4}. Thus, it is difficult for the MUSIC algorithm to explore such orthogonality when noise is scarce, leading to the slow speed of the performance that increases with SNR. If the DoA estimation performance is expected to be further improved, substituting the MUSIC algorithm with its improved versions \cite{ref38,ref39,ref40} is a preferable choice.

\begin{figure*}[!tb]
\vspace{0mm}
\centering
\subfloat[]{\includegraphics[width=2.15in]{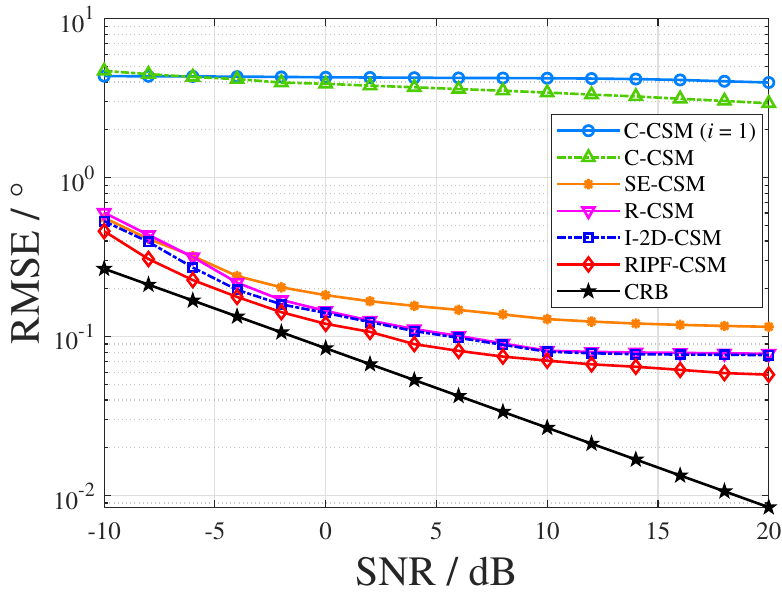}
\label{fig3a}}
\hfil
\subfloat[]{\includegraphics[width=2.15in]{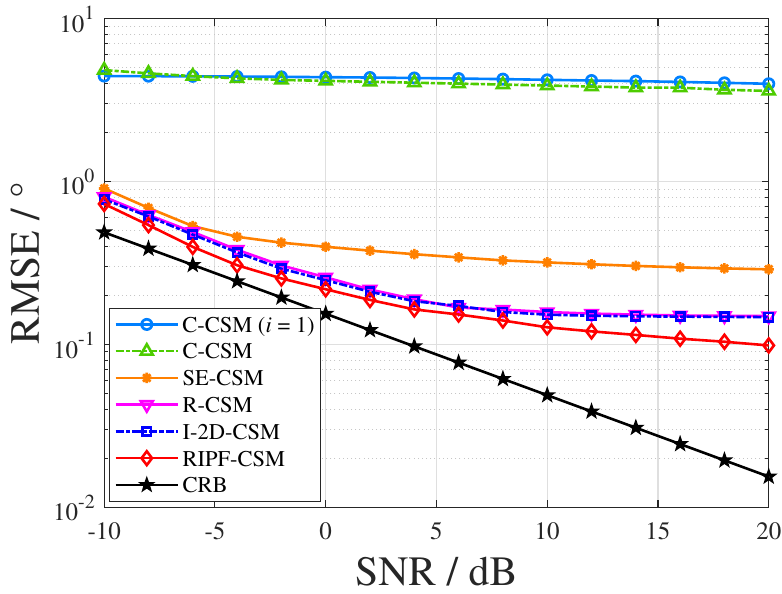}
\label{fig3b}}
\hfil
\subfloat[]{\includegraphics[width=2.15in]{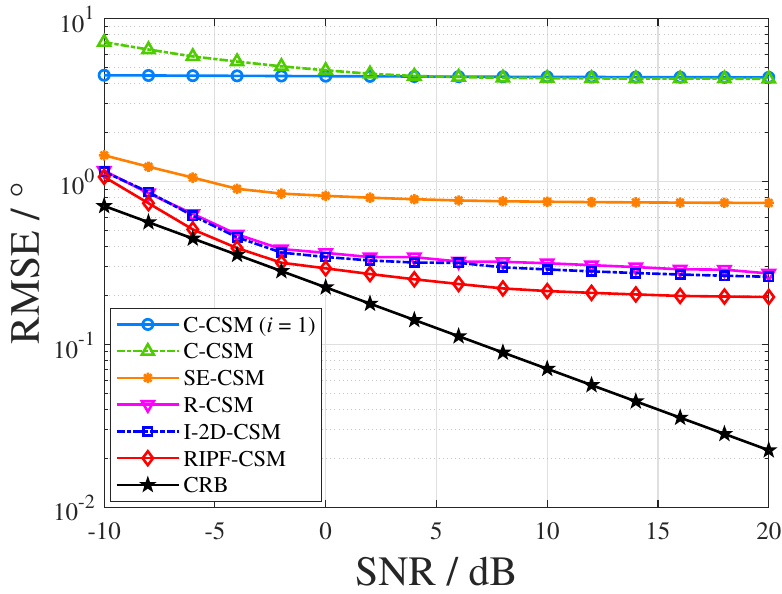}
\label{fig3c}}
\hfil
\subfloat[]{\includegraphics[width=2.15in]{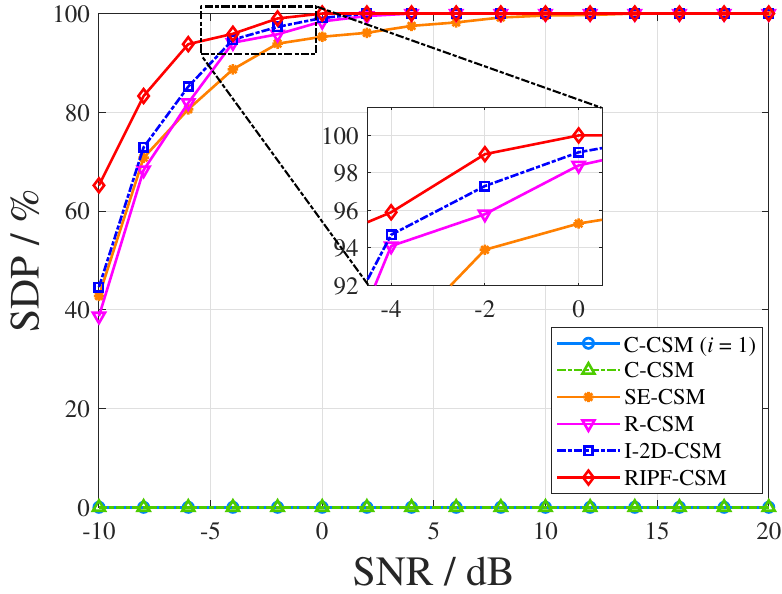}
\label{fig3d}}
\hfil
\subfloat[]{\includegraphics[width=2.15in]{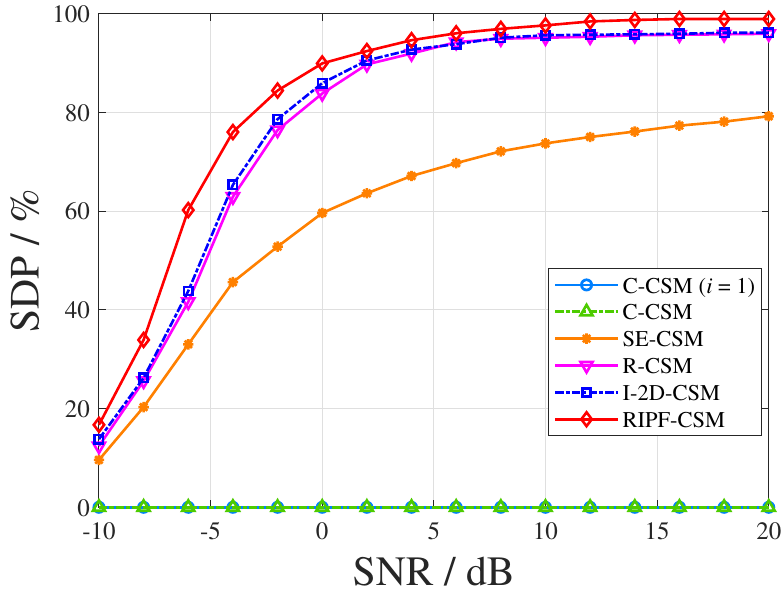}
\label{fig3e}}
\hfil
\subfloat[]{\includegraphics[width=2.15in]{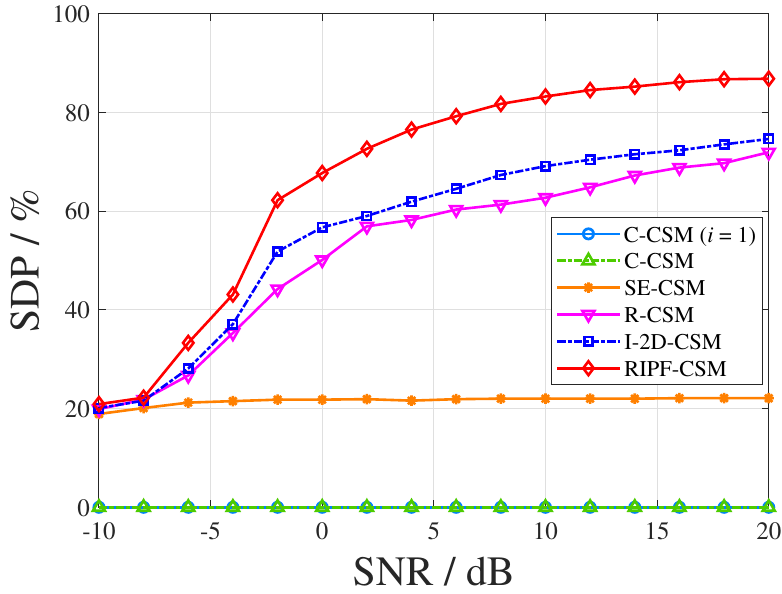}
\label{fig3f}}
\caption{RMSEs and SDPs of estimation results in $\text{SNR}\in[-10\text{dB},20\text{dB}]$. (a) and (d): $N=1$; (b) and (e): $N=2$; (c) and (f): $N=3$.}
\label{fig3}
\vspace{-2mm}
\end{figure*}

\subsection{RMSE and Running Time Versus the Average Error of Pre-estimated DoAs}

\begin{figure*}[!tb]
\centering
\subfloat[]{\includegraphics[width=2.15in]{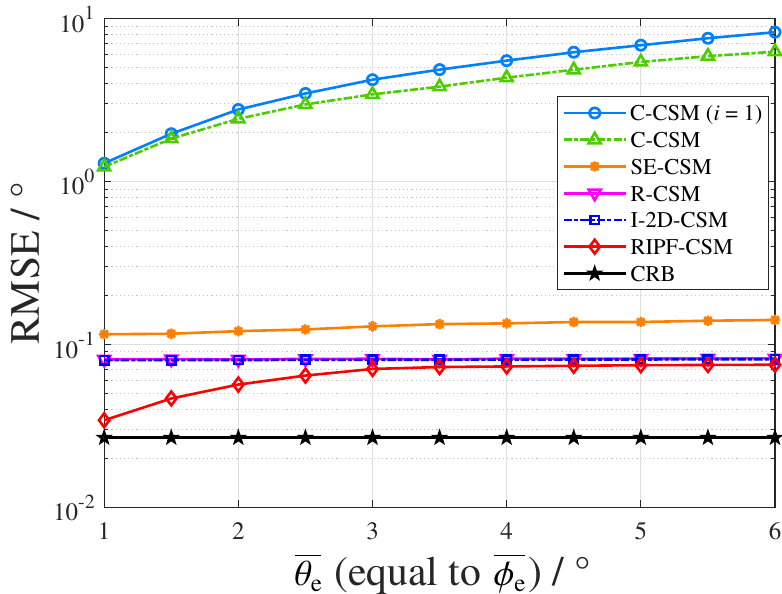}
\label{fig4a}}
\hfil
\subfloat[]{\includegraphics[width=2.15in]{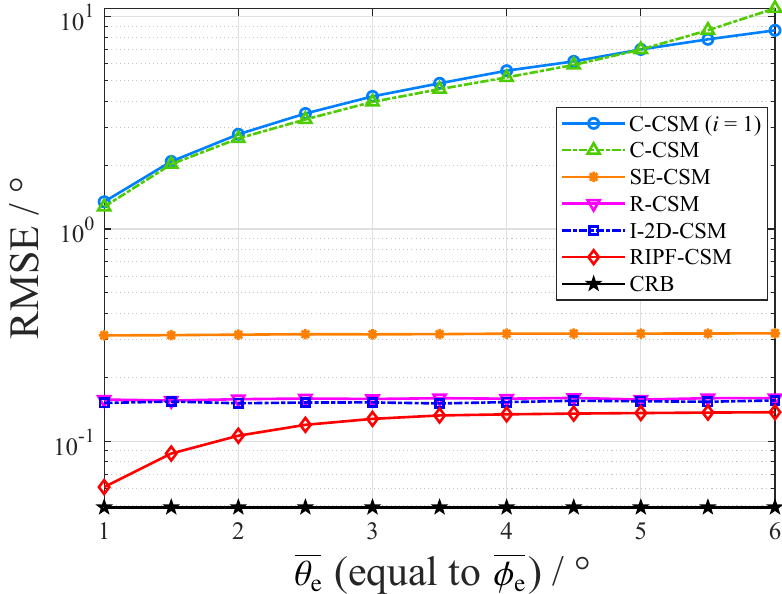}
\label{fig4b}}
\hfil
\subfloat[]{\includegraphics[width=2.15in]{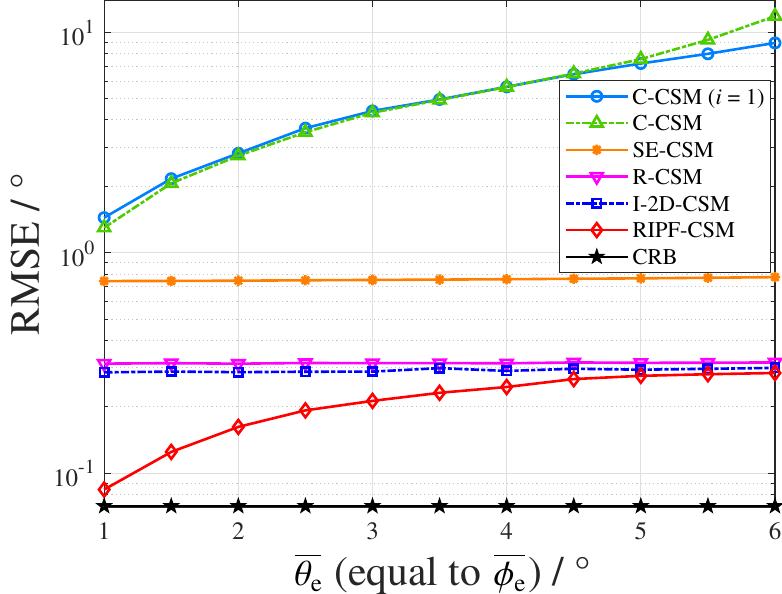}
\label{fig4c}}
\hfil
\subfloat[]{\includegraphics[width=2.15in]{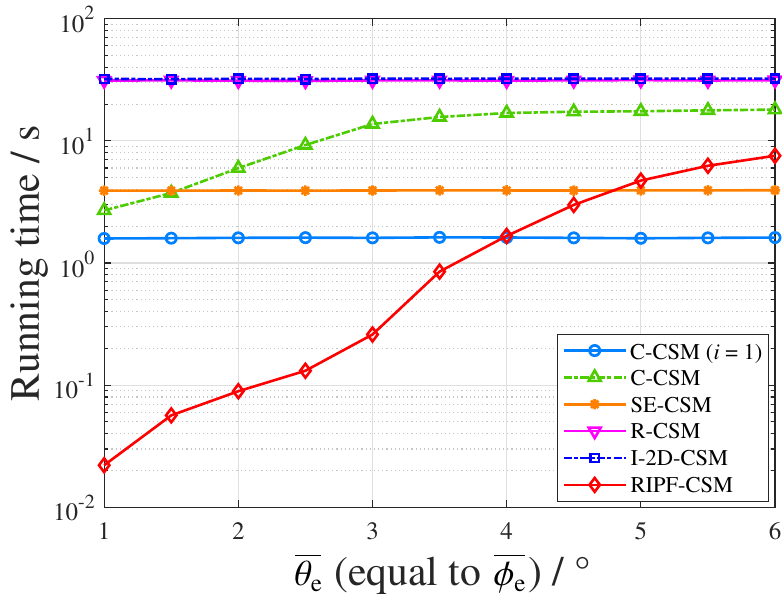}
\label{fig4d}}
\hfil
\subfloat[]{\includegraphics[width=2.15in]{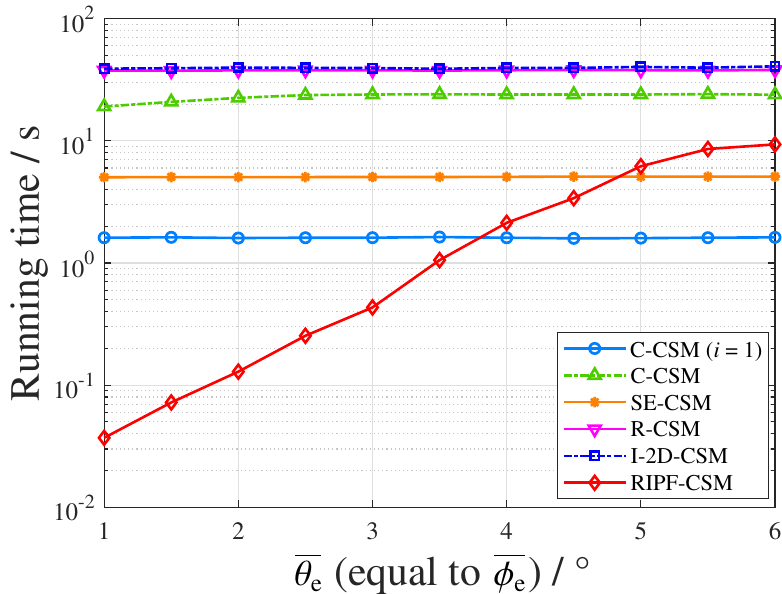}
\label{fig4e}}
\hfil
\subfloat[]{\includegraphics[width=2.15in]{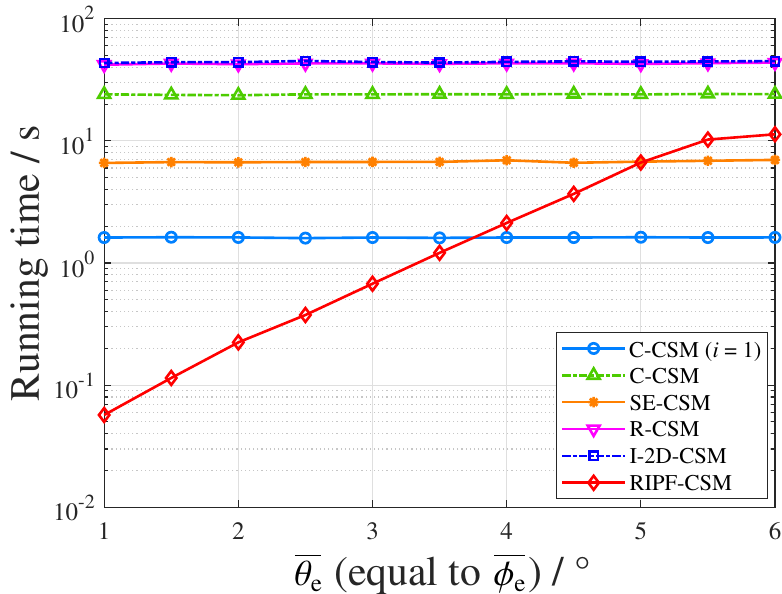}
\label{fig4f}}
\caption{RMSEs of estimation results and running times in $\overline{\theta_\text{e}},\overline{\phi_\text{e}}\in[1^\circ,6^\circ]$. (a) and (d): $N=1$; (b) and (e): $N=2$; (c) and (f): $N=3$.}
\label{fig4}
\vspace{-2mm}
\end{figure*}

Compared with SDP, RMSE can better reflect the error of estimation results according to (34) and (36). Thus, RMSE is chosen to evaluate the estimation accuracy in this subsection. The RMSEs and the running times of different methods at different $\overline{\theta_\text{e}}$, $\overline{\phi_\text{e}}$ and $N$ are shown in Fig. 4, as well as the RMSEs of the average CRBs. In Fig. 4, Figs. 4 (a)-(c) and Figs. 4 (d)-(f) respectively shows the RMSEs and the running times with $\overline{{{\theta }_\text{e}}},\overline{{{\phi }_\text{e}}}\in [{{1}^{\circ }},{{6}^{\circ }}]$, where $N=1,2,3$ are respectively applied to Figs. 4 (a) and (d), Figs. 4 (b) and (e), Figs. 4 (c) and (f).

It can be observed from Figs. 4 (a)-(c) that for any $\overline{\theta_\text{e}}$, $\overline{\phi_\text{e}}$ and $N$ in the simulations, RIPF-CSM exhibits better performance when compared with the benchmark methods. Additionally, the RMSE of RIPF-CSM is closer to CRB with the decrease of $\overline{{{\theta }_\text{e}}}$ and $\overline{{{\phi }_\text{e}}}$. These phenomenons demonstrate that the proposed robustness intervals can make full use of the pre-estimated DoAs and reveals the robustness of RIPF-CSM to different errors of pre-estimated DoAs. In contrast, Figs. 4 (a)-(c) show that the RMSEs of SE-CSM, R-CSM and I-2D-CSM almost keep constant for any $\overline{{{\theta }_\text{e}}}$ and $\overline{{{\phi }_\text{e}}}$. It is due to the reason that most of their focusing angles are far away from the actual DoAs, which makes the focusing process be insensitive to the accuracy of the pre-estimated DoAs. Furthermore, since the robustness intervals of RIPF-CSM become larger when $\overline{{{\theta }_\text{e}}}$ and $\overline{{{\phi }_\text{e}}}$ increase, the performance degrades to almost the same as that of R-CSM and I-2D-CSM as shown in Figs. 4 (a)-(c).

Figs. 4 (d)-(f) show that the running time of RIPF-CSM decreases approximately in the square order with the decrease of $\overline{{{\theta }_\text{e}}}$ and $\overline{{{\phi }_\text{e}}}$. Such phenomenon coincides with the expression in Table 1 that the radii of the robustness intervals as well as the number of iterations decrease when the pre-estimated DoAs are more accurate. On the contrary, it is observed that generally the running times of the benchmark methods cannot benefit from the pre-estimated DoAs even if they are accurate. It is due to the fact that C-CSM and SE-CSM are difficult to converge and often terminates at the upper limit of iteration number, and the initial robustness intervals of R-CSM and I-2D-CSM are so large that $\overline{{{\theta }_\text{e}}}$ and $\overline{{{\phi }_\text{e}}}$ can slightly affect the estimation accuracy and thus the convergence speed. In Fig. 4 (d), although C-CSM can converge when the pre-estimated DoAs are accurate enough, the high complexity from its focusing process and the MUSIC algorithm leads to longer running time than RIPF-CSM. When $\overline{{{\theta }_\text{e}}}$ and $\overline{{{\phi }_\text{e}}}$ are not greater than $3.5^\circ$, the running time of RIPF-CSM is shorter than other methods as illustrated in Figs. 4 (d)-(f), which exhibits the high efficiency contributed by the integration of the less candidate frequency points, the robustness intervals and the range-shrunk spatial spectrum. According to the above analysis, RIPF-CSM achieves a better trade-off between performance and complexity than the other methods when $\overline{{{\theta }_\text{e}}},\overline{{{\phi }_\text{e}}}\in [{{1}^{\circ }},{{6}^{\circ }}]$ that is typical in practical scenarios \cite{ref18,ref19,ref20,ref21}.

\subsection{Running Time Versus the Angle Sampling Steps and the Number of Frequency Points}

\begin{figure*}[!tb]
\centering
\subfloat[]{\includegraphics[width=2.15in]{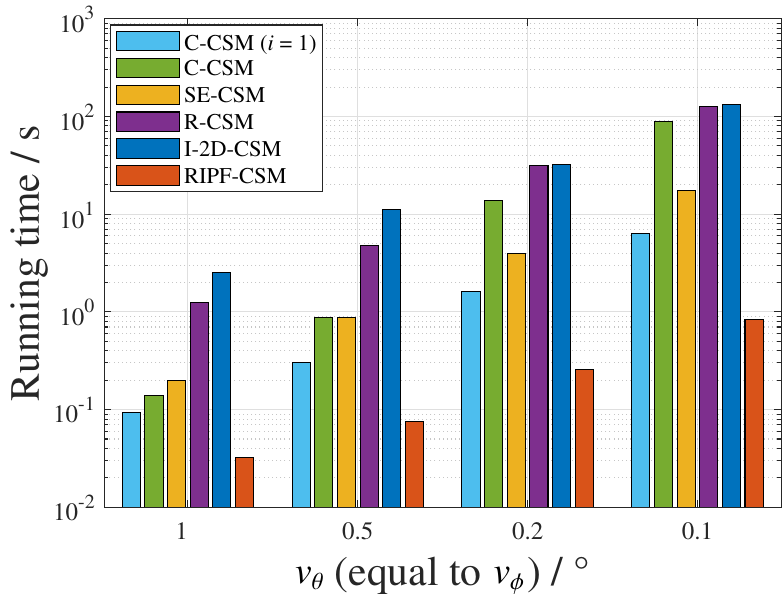}
\label{fig5a}}
\hfil
\subfloat[]{\includegraphics[width=2.15in]{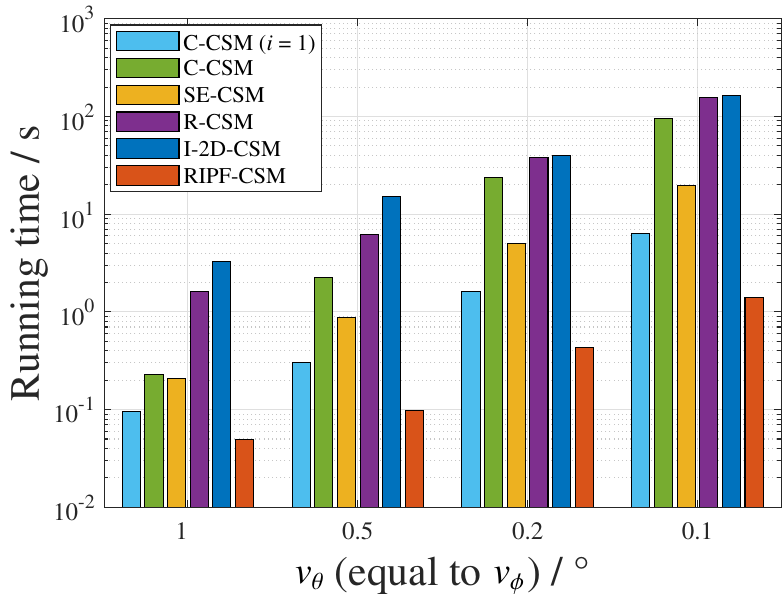}
\label{fig5b}}
\hfil
\subfloat[]{\includegraphics[width=2.15in]{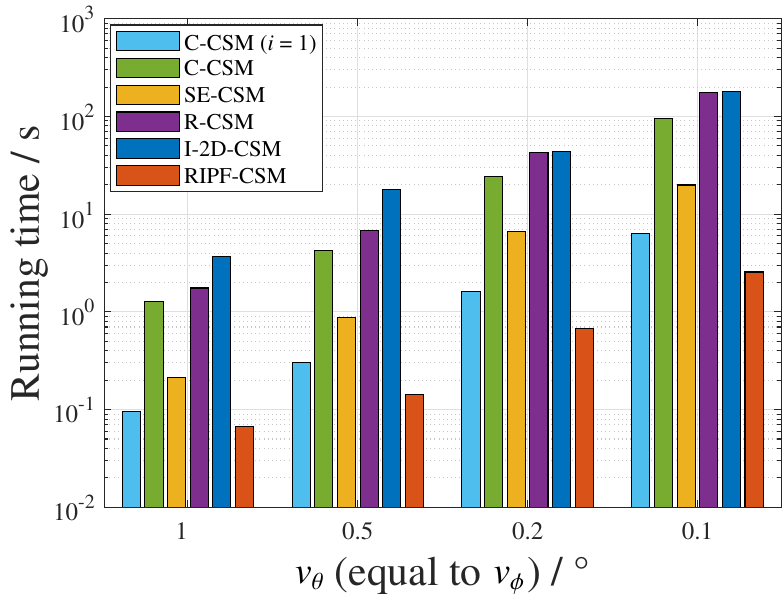}
\label{fig5c}}
\hfil
\subfloat[]{\includegraphics[width=2.15in]{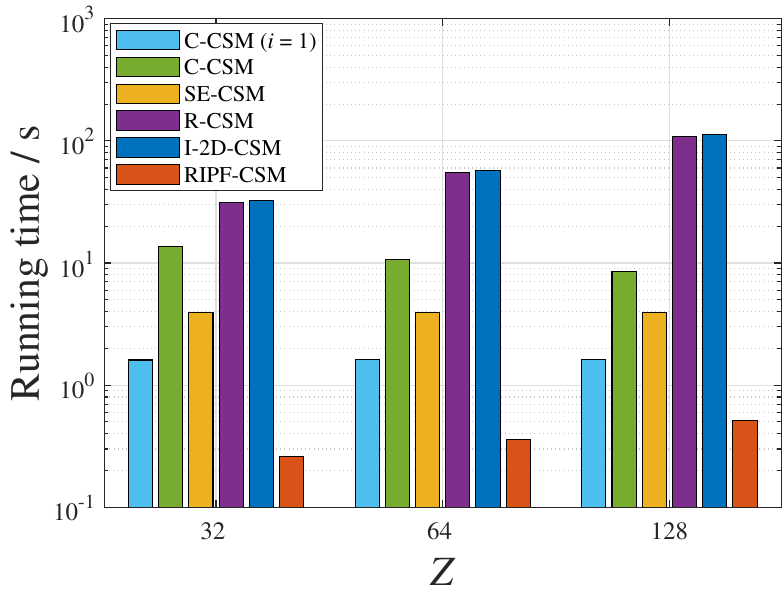}
\label{fig5d}}
\hfil
\subfloat[]{\includegraphics[width=2.15in]{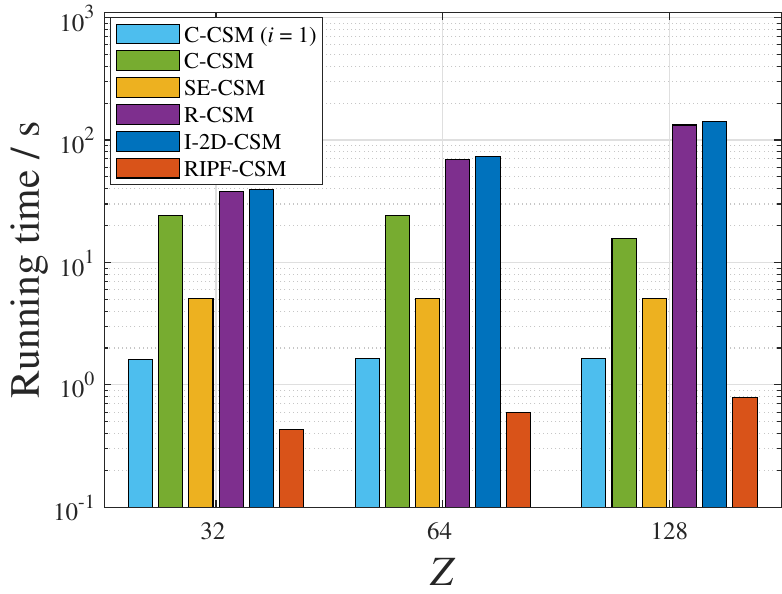}
\label{fig5e}}
\hfil
\subfloat[]{\includegraphics[width=2.15in]{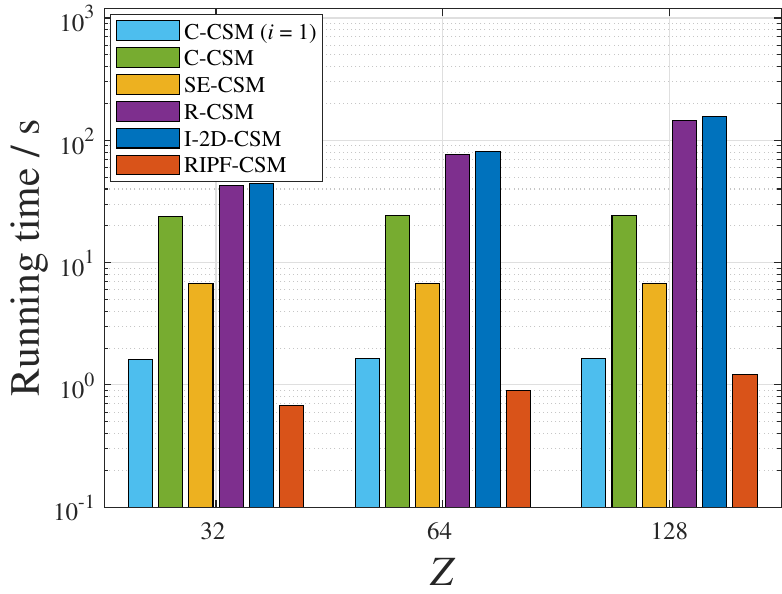}
\label{fig5f}}
\caption{Running times in $v_\theta,v_\phi\in\{1^\circ,0.5^\circ,0.2^\circ,0.1^\circ\}$ and $Z\in\{32,64,128\}$. (a) and (d): $N=1$; (b) and (e): $N=2$; (c) and (f): $N=3$.}
\label{fig5}
\vspace{-2mm}
\end{figure*}

Note from Table 1 that the angle sampling steps $v_\theta$ and $v_\phi$, and the number $Z$ of frequency points severely affect the computational complexity of each method. To explicitly reveal such dependence, the running times of RIPF-CSM and the benchmark methods are respectively measured with different angle sampling steps and the number of frequency points. The results with $N=1,2,3$ are shown in Fig. 5.

It is observed from Figs. 5 (a)-(c) that the running times of different methods increase with the reduction of $v_\theta$ and $v_\phi$, which indicates higher angle resolution requires more computational complexity. In addition, Figs. 5 (d)-(f) shows that the running times of most methods increase with $Z$ except for the C-CSM with $N=1,2$. Generally, the running speed is negatively related with $Z$, but the increase of $Z$ helps improve the accuracy of DoA estimation in each iteration so that the number of iterations required for convergence can be reduced. As an example, C-CSM converges before reaching the upper limit of the number of iterations when $N=1$ with $Z=32,64,128$ or $N=2$ with $Z=128$, and thus the corresponding running times is shorter than the cases with smaller $Z$.

Furthermore, as illustrated in Fig. 5, RIPF-CSM runs faster than the other methods for any $v_\theta$, $v_\phi$ or $Z$, which demonstrates the high efficiency of RIPF-CSM. Besides, it is shown in Fig. 5 (a)-(c) that the differences of the running times between RIPF-CSM and the benchmark methods increase with the decrease of the angle sampling steps. Therefore, such phenomenon enables RIPF-CSM to be more advantageous for the applications that require the DoA estimation with higher resolution when compared with the benchmark methods.

\section{Conclusions}
In this paper, RIPF-CSM is proposed for UCAs to iteratively estimate the 2D DoAs of wideband signals. In RIPF-CSM, the focusing process with the proposed robustness intervals achieves high performance, and the preferable efficiency of this method is obtained by shrinking the range of the MUSIC spatial spectrum to the regions constrained by the robustness intervals and reducing the number of the candidate frequency points which increases flexibly with iterations. Moreover, the reliability of RIPF-CSM is improved by adjusting the key parameters with the estimated DoAs in the previous iterations. Besides, this paper derives a parameter constraint that guarantees the computational complexity of RIPF-CSM lower than that of any benchmark method in typical scenarios. As a theoretical benchmark of the performance of RIPF-CSM, the CRB of the 2D DoA estimation using a UCA is derived with the consideration of multiple wideband signal sources and multiple snapshots. Based on the derived CRB, this paper provides the guidelines for practical application of the DoA estimation in the corresponding scenario. The simulation results reveal that RIPF-CSM outperforms the benchmark methods, and its performance is closer to CRB. Furthermore, RIPF-CSM is efficient under a variety of simulation settings. Its advantage in terms of the computational complexity is obvious when the pre-estimated DoAs are accurate enough or the DoA estimation with high angle resolution is required.

\appendix

\section{Dimension Extension of the Benchmark Methods}
SE-CSM in \cite{ref19}, R-CSM in \cite{ref22} and I-2D-CSM in \cite{ref23} are only designed for 1D DoA estimation. For fair comparison with RIPF-CSM, they are extended to the versions for 2D DoA estimation in this appendix, where the key of extension is the reselection of the focusing angles. The details of the extension are explained as follows.

1) \textit{SE-CSM}:

In \cite{ref19}, the focusing angles except for the pre-estimated DoAs, i.e., the extra focusing angles, in the first iteration are set as $\hat{\theta }_{n}^{(0)}\pm 0.25\text{BW}_\theta$, where $\text{BW}_\theta$ denotes the vertical $3$dB beamwidth, and $n=1,2,\dots,{{\hat{N}}_{i-1}}$. In the subsequent iterations, these focusing angles are set to be $\hat{\theta }_{n}^{(i-1)}\pm 0.125\text{BW}_\theta$.

Following this idea, in the extended SE-CSM for 2D DoA estimation, the extra focusing angles in the first iteration are set as $(\hat\theta_n^{(0)}\pm 0.25\text{BW}_\theta ,\hat\phi_n^{(0)}\pm 0.25\text{BW}_\phi)$, and the ones in subsequent iterations are set as $(\hat\theta_n^{(i-1)}\pm 0.125\text{BW}_\theta,\hat\phi_n^{(i-1)}\pm 0.125\text{BW}_\phi)$. Since the beamwidth of UCA varies with different elevation angles \cite{ref41}, $\text{BW}_\theta$ and $\text{BW}_\phi$ respectively stand for the vertical and horizontal $3$dB beamwidth of the quiescent array pattern \cite{ref42}, i.e., the direction of the beam is at $(0^\circ,0^\circ)$.

2) \textit{R-CSM}:

The focusing angles in the $i$-th iteration of the R-CSM in \cite{ref22} are obtained by sampling the elevation robustness interval $\mathbb{G}{{_{n}^{\theta }}^{\prime }}(i)$ with $v_\theta$, which is expressed as:
\begin{equation}
\mathbb{G}{{_{n}^{\theta }}^{\prime }}(i)=\bigg[\arcsin\Big( \max \big[ 0,\sin\hat{\theta }_{n}^{(i-1)}-\frac{1}{2{{i}^{p}}} \big]\Big),\arcsin\Big( \min \big[ \sin\hat{\theta }_{n}^{(i-1)}+\frac{1}{2{{i}^{p}}},1 \big]\Big) \bigg]\text{,}
\end{equation}
where $n=1,2,\dots,{{\hat{N}}_{i-1}}$, $p=2$, and its radius $R{{_{n}^{\theta }}^{\prime }}(i)$ is the half of the length of $\mathbb{G}{{_{n}^{\theta }}^{\prime }}(i)$. In (A.1), $1/(2i^p)$ can be considered as the radius of the interval mapped from $\mathbb{G}{{_{n}^{\theta }}^{\prime }}(i)$ by the sine function whose codomain is $[0,1]$ with $\theta\in[0^\circ,90^\circ]$. However, this idea cannot be applied to the radius for an azimuth interval mapped by the sine function, since $\phi\in[0^\circ,360^\circ)$ and thus the mapping is no longer injective. In this sense, the radius of the azimuth robustness interval is set in angle domain, i.e., $R{{_{n}^{\phi }}^{\prime }}(i)={{360}^{\circ }}/(2i^p)$. With regarding $\hat{\phi}_{n}^{(i-1)}$ as the center of the azimuth robustness interval $\mathbb{G}{{_{n}^{\phi }}^{\prime }}(i)$, this interval can be obtained by replacing $R_{n}^{\phi }(i)$ in (16) with $R{{_{n}^{\phi }}^{\prime }}(i)$.

3) \textit{I-2D-CSM}:

The elevation robustness interval of I-2D-CSM in \cite{ref23} is similar to that of R-CSM, from which the focusing angles are sampled with $v_\theta$. This interval is expressed as:
\begin{equation}
\mathbb{G}{{_{n}^{\theta }}^{\prime\prime }}(i)=\bigg[\arcsin\Big(\max \big[0,\sin\hat{\theta }_{n}^{(i-1)}-\frac{1}{2\min[i,i_\text{s}^\theta]^2} \big]\Big),\arcsin\Big(\min \big[\sin\hat{\theta}_{n}^{(i-1)}+\frac{1}{2\min[i,i_\text{s}^\theta]^2},1 \big]\Big)\bigg]\text{,}
\end{equation}
where $n=1,2,\dots,{{\hat{N}}_{i-1}}$, $i_{\text{s}}^{\theta }=\rho {{v}_{\theta }}/1^\circ$, $\rho \in (1,3]$, and its radius $R{{_{n}^{\theta }}^{\prime\prime }}(i)$ is the half of the length of $\mathbb{G}{{_{n}^{\theta }}^{\prime\prime}}(i)$. In this paper, $\rho$ is set as $2$. By imitating the extension of the above R-CSM, the radius of the azimuth robustness interval in the $i$-th iteration is set as $R{{_{n}^{\phi }}^{\prime\prime }}(i)={{360}^{\circ }}/(2\min [i,i_\text{s}^\phi]^{2})$ where $i_{\text{s}}^{\phi }=\rho {{v}_{\phi }}/1^\circ$. Set $\hat{\phi}_{n}^{(i-1)}$ as the center of the azimuth robustness interval $\mathbb{G}{{_{n}^{\phi }}^{\prime\prime }}(i)$, and then this interval can be obtained by replacing $R_{n}^{\phi }(i)$ in (16) with $R{{_{n}^{\phi }}^{\prime\prime }}(i)$.

\section{Parameter Constraint Derivation for RIPF-CSM with Lower Computational Complexity than C-CSM in a Single Iteration}
For a single iteration, there is a constraint for the parameters mentioned in Table 1 to guarantee the lower computational complexity of RIPF-CSM when compared with C-CSM in common scenarios, which is discussed as follows.

The constraint is derived on the premise of the following two assumptions, which are generally the cases in practice. First, the changes of the results between adjacent iterations are large only in the first few iterations, and the changes are smaller in the subsequent iterations. Statistically, these small changes appear more frequently than the large ones according to \cite{ref22} and the simulations. Based on these arguments, it is reasonable to assume $d_{\overline{\delta }}^{(i)}$ in (9), $i= 1,2,\dots,I_\text{c}$, follows a certain exponential distribution. Second, $d_\theta$ and $d_\phi$ can represent the approximate errors of the pre-estimated elevation and azimuth. Considering the estimation results generally approach the actual DoAs with iterations, and the final estimation errors are close to zero,  $d_{\overline{\delta }}^{(i)}$ is assumed to satisfy:
\begin{equation}
\mathcal{E}\bigg[\sum\limits_{i=1}^{{{I}_\text{c}}}{d_{\bar\delta}^{(i)}}\bigg]=\frac{{{d}_{\theta }}+{{d}_{\phi }}}{2}\text{,}
\end{equation}
where $I_\text{c}$ denotes the number of iterations at the termination. Thus, $\mathcal{E}[d_{\overline{\delta }}^{( i )}]=(d_\theta+d_\phi)/(2I_\text{c})$ holds. In addition, due to the cumulative distribution function of exponential distribution, there exists:
\begin{equation}
\mathcal{P}\bigg[\sum\limits_{i^\prime=1}^{i}{d_{\bar\delta}^{(i^\prime)}}<\frac{\ln 100(d_\theta+d_\phi)i}{2I_\text{c}}\bigg]\le (99\%)^{i}\text{,}
\end{equation}
where $i\le I_\text{c}$ and $I_\text{c}$ is generally less than $5$ as mentioned in Subsection 3.3. With such high probability in (B.2), the following inequality approximately holds:
\begin{equation}
\sum\limits_{i^\prime=1}^{i}{d_{\bar\delta}^{(i^\prime)}}<\frac{\ln 100(d_\theta+d_\phi)}{2}\text{.}
\end{equation}

Based on the above discussions, the derivation of the parameter constraint is given as follows. By combining (9), (10) and (B.3), there exists:
\begin{equation}
Z_{\text{in}}^{(i)}\le 1+\frac{\ln 100({{d}_{\theta }}+{{d}_{\phi }})\left[( {{d}_{\theta }}+{{d}_{\phi }})I+2Z \right]}{4{{I}_\text{c}}I}\text{.}
\end{equation}
It is known from $\hat{\theta }_{n}^{( i )}\in[0^\circ,90^\circ]$ that $\max[(b-\sin\hat{\theta }_{n}^{( i )})(b-\cos\hat{\theta }_{n}^{( i )})]=b(b-1)$. According to this equation, $i\ge1$, (11), (12), $\mathcal{E}[d_{\overline{\delta }}^{(i)}]$ and the cumulative distribution function of exponential distribution, $\sum\limits_{n=1}^{\hat{N}_{i}}{R_{n}^{\theta }( i )R_{n}^{\phi }( i )}$ satisfies:
\begin{equation}
\mathcal{P}\bigg[ \sum\limits_{n=1}^{\hat{N}_{i}}{R_{n}^{\theta }(i)R_{n}^{\phi }(i)}<\Big(\frac{\ln 100}{2I_\text{c}}\Big)^{2}{\hat{N}_{i}}\overline{\theta_\text{e}}\hspace{0.4mm}\overline{\phi_\text{e}}{{( {{d}_{\theta }}+{{d}_{\phi }})}^{2}}b(b-1) \bigg]\le 99\%\text{.}
\end{equation}
On the basis of Table 1, when the computational complexity of RIPF-CSM is lower than that of C-CSM in a single iteration, the following inequality is satisfied:
\begin{equation}
Z_{\text{in}}^{(i)}\Bigg[M+8{{K}_{f}}+16\sum\limits_{n=1}^{\hat{N}_{i}}{R_{n}^{\theta }(i)R_{n}^{\phi }(i)}/({{v}_{\theta }}{{v}_{\phi }})\Bigg]<Z(M+4{\hat{N}_{i}}+8{{K}_{f}})+4{{L}_{\theta }}{{L}_{\phi }}\text{.}
\end{equation}
By substituting the right expression of (B.4) and the right expression in the $\mathcal{P}[\cdot]$ of (B.5) into $Z_{\text{in}}^{( i )}$ and $\sum\limits_{n=1}^{\hat{N}_{i}}{R_{n}^{\theta }( i )R_{n}^{\phi }( i )}$ in (B.6) respectively, the probability that the computational complexity of RIPF-CSM in a single iteration is lower than that of C-CSM is high enough. The result of the above substitution is the desired parameter constraint, which can be written as:
\begin{equation}
\begin{aligned}
&\Big[4I^2+\ln 100(d_\theta+d_\phi)\big[(d_\theta+d_\phi)I+2Z\big]\Big]\times\Big[M+8K_f+\frac{(2\ln 100)^2}{v_\theta v_\phi I^2}N\overline{\theta_\text{e}}\hspace{0.4mm}\overline{\phi_\text{e}}(d_\theta+d_\phi)^2b(b-1)\Big]\\
<&\hspace{1mm}4I^2[Z(M+4N+8K_f)+4L_\theta L_\phi]\text{,}
\end{aligned}
\end{equation}
where $\hat{N}_{i}$ and $I_\text{c}$ are replaced by $N$ and $I$ for generality, respectively. This parameter constraint is easy to be satisfied by most of the simulation parameters in Section 5, which are common in practical scenarios. For instance, the default simulation parameters with $N\in[1,3]$ satisfy this constraint. With such high probability of (B.6) that holds, it is reasonable to regard that RIPF-CSM enjoys the lower computational complexity than that of C-CSM in a single iteration.

\section{Derivation of the Fisher Information Matrix}

\noindent{\bf Lemma 1.} {\it Suppose $k_1,k_2\in\{1,2,\dots,K_f\}$ and $z_1,z_2\in\{1,2,\dots,Z\}$, then:}
\begin{equation}
\mathcal{E}\left[ {{\mathbf{w}}^{\text{H}}}( {{k}_{1}},{{f}_{{{z}_{1}}}})\mathbf{w}( {{k}_{1}},{{f}_{{{z}_{1}}}}){{\mathbf{w}}^{\text{H}}}( {{k}_{2}},{{f}_{{{z}_{2}}}})\mathbf{w}( {{k}_{2}},{{f}_{{{z}_{2}}}}) \right]=
\begin{cases}
M\left( M+1 \right){{Z}^{2}}{{\sigma }^{4}}, & \text{if }{{k}_{1}}={{k}_{2}}\text{ and }{{z}_{1}}={{z}_{2}}\text{,} \\
{{M}^{2}}{{Z}^{2}}{{\sigma }^{4}}, & \text{otherwise,}
\end{cases}
\end{equation}
\begin{equation}
\mathcal{E}\left[ {{\mathbf{w}}^{\text{H}}}( {{k}_{1}},{{f}_{{{z}_{1}}}} )\mathbf{w}( {{k}_{1}},{{f}_{{{z}_{1}}}}){{\mathbf{w}}^{\text{T}}}( {{k}_{2}},{{f}_{{{z}_{2}}}}) \right]=0\text{.}
\end{equation}
{\bf Proof.} This lemma can be easily obtained based on the correlation properties of uncorrelated Gaussian noise.

\noindent\textbf{The derivation of the Fisher information matrix:}

Denote $d_{\sigma^{2}}=\partial\ln P/\partial\sigma^{2}$, $\mathbf{d}_{\mathbf{\hat{s}}_f}^{k,z}=\partial\ln P/\partial\mathbf{\hat{s}}_f(k,f_z)$, $\mathbf{d}_{\mathbf{\check{s}}_f}^{k,z}=\partial\ln P/\partial\mathbf{\check{s}}_f(k,f_z)$ and $\mathbf{d}_{\boldsymbol{\zeta}}=\partial\ln P/\partial\boldsymbol{\zeta}$, which are respectively calculated as:
\begin{equation}
d_{\sigma^{2}}=-\frac{{{K}_{f}}MZ}{{{\sigma }^{2}}}+\frac{1}{Z{{\sigma }^{4}}}\sum\limits_{k=1}^{{{K}_{f}}}{\sum\limits_{z=1}^{Z}{{{\mathbf{w}}^{\text{H}}}( k,{{f}_{z}})\mathbf{w}( k,{{f}_{z}})}}\text{,}
\end{equation}
\begin{equation}
\mathbf{d}_{\mathbf{\hat{s}}_f}^{k,z}=\frac{2}{Z\sigma^2}\operatorname{Re}\left[ {{\mathbf{A}}^{\text{H}}}( {{f}_{z}})\mathbf{w}( k,{{f}_{z}} ) \right]\text{,}
\end{equation}
\begin{equation}
\mathbf{d}_{\mathbf{\check{s}}_f}^{k,z}=\frac{2}{Z\sigma ^2}\operatorname{Im}\left[ {{\mathbf{A}}^{\text{H}}}( {{f}_{z}})\mathbf{w}( k,{{f}_{z}} ) \right]\text{,}
\end{equation}
\begin{equation}
\mathbf{d}_{\boldsymbol{\zeta}}=\frac{2}{Z{{\sigma }^{2}}}\sum\limits_{k=1}^{{{K}_{f}}}{\sum\limits_{z=1}^{Z}{\operatorname{Re}\left[ {{\mathbf{\Xi }}^{\text{H}}}( k,{{f}_{z}}){{\mathbf{D}}^{\text{H}}}( {{f}_{z}})\mathbf{w}( k,{{f}_{z}}) \right]}}\text{.}
\end{equation}
Based on (C.3)-(C.6) and Lemma 1, the following equations hold:
\begin{equation}
\mathcal{E}\left[d_{\sigma^{2}}^{2} \right]=\frac{{{K}_{f}}MZ}{{{\sigma }^{4}}}\text{,}
\end{equation}
\begin{equation}
\mathcal{E}\left[d_{\sigma^{2}}\mathbf{d}_{\mathbf{\hat{s}}_f}^{k,z} \right]=\mathcal{E}\left[ d_{\sigma^{2}}\mathbf{d}_{\mathbf{\check{s}}_f}^{k,z} \right]=\mathcal{E}\left[d_{\sigma^{2}} \mathbf{d}_{\boldsymbol{\zeta}} \right]=0\text{,}
\end{equation}
\begin{equation}
\mathcal{E}\left[ \mathbf{d}_{\mathbf{\hat{s}}_f}^{k,z}\mathbf{d}_{\boldsymbol{\zeta}}^{\text{T}} \right]=\frac{2}{Z{{\sigma }^{2}}}\operatorname{Re}\left[ {{\mathbf{A}}^{\text{H}}}({{f}_{z}})\mathbf{D}({{f}_{z}})\mathbf{\Xi }( k,{{f}_{z}} ) \right]\text{,}
\end{equation}
\begin{equation}
\mathcal{E}\left[\mathbf{d}_{\mathbf{\check{s}}_f}^{k,z}\mathbf{d}_{\boldsymbol{\zeta}}^{\text{T}} \right]=\frac{2}{Z{{\sigma }^{2}}}\operatorname{Im}\left[ {{\mathbf{A}}^{\text{H}}}({{f}_{z}})\mathbf{D}({{f}_{z}})\mathbf{\Xi }(k,f_z) \right]\text{,}
\end{equation}
\begin{equation}
\mathcal{E}\left[ \mathbf{d}_{\boldsymbol{\zeta}}\mathbf{d}_{\boldsymbol{\zeta}}^{\text{T}} \right]=\frac{2}{Z{{\sigma }^{2}}}\sum\limits_{k=1}^{{{K}_{f}}}{\sum\limits_{z=1}^{Z}{\operatorname{Re}\left[ {{\mathbf{\Xi }}^{\text{H}}}(k,{{f}_{z}}){{\mathbf{D}}^{\text{H}}}({{f}_{z}})\mathbf{D}({{f}_{z}})\mathbf{\Xi}(k,{{f}_{z}}) \right]}}\text{,}
\end{equation}
\begin{equation}
\mathcal{E}\left[ \mathbf{d}_{\mathbf{\check{s}}_f}^{k,z}{\mathbf{d}_{\mathbf{\check{s}}_f}^{k,z}}^{\text{T}} \right]=\mathcal{E}\left[ \mathbf{d}_{\mathbf{\hat{s}}_f}^{k,z}{\mathbf{d}_{\mathbf{\hat{s}}_f}^{k,z}}^{\text{T}} \right]=
\begin{cases}
\frac{2}{Z{{\sigma }^{2}}}\operatorname{Re}\left[ {{\mathbf{A}}^{\text{H}}}( {{f}_{{{z}_{1}}}})\mathbf{A}( {{f}_{{{z}_{2}}}}) \right], & \text{if } {{k}_{1}}={{k}_{2}}\text{ and }{{z}_{1}}={{z}_{2}}\text{,} \\
0, & \text{otherwise,}
\end{cases}
\end{equation}
\begin{equation}
\mathcal{E}\left[ \mathbf{d}_{\mathbf{\hat{s}}_f}^{k,z}{\mathbf{d}_{\mathbf{\check{s}}_f}^{k,z}}^{\text{T}} \right]=-\mathcal{E}\left[ \mathbf{d}_{\mathbf{\check{s}}_f}^{k,z}{\mathbf{d}_{\mathbf{\hat{s}}_f}^{k,z}}^{\text{T}} \right]=
\begin{cases}
-\frac{2}{Z{{\sigma }^{2}}}\operatorname{Im}\left[ {{\mathbf{A}}^{\text{H}}}( {{f}_{{{z}_{1}}}})\mathbf{A}( {{f}_{{{z}_{2}}}}) \right], & \text{if } {{k}_{1}}={{k}_{2}}\text{ and }{{z}_{1}}={{z}_{2}}\text{,} \\
0, & \text{otherwise,}
\end{cases}
\end{equation}
where $k,k_1,k_2\in\{1,2,\dots,K_f\}$ and $z,z_1,z_2\in\{1,2,\dots,Z\}$. Then, (20) can be obtained by substituting (C.7)-(C.13) into $\mathbf{F}=\mathcal{E}[\boldsymbol{\psi }{{\boldsymbol{\psi}}^{\text{T}}}]$ which is mentioned in Subsection 4.1.

\section*{Acknowledgement}

This work was supported in part by the National Natural Science Foundation of China under Grant 62371053, 61871050, and the US National Science Foundation under Grant 2136202.

\end{document}